\documentclass[12pt,draft]{article}

\usepackage[english]{babel}

\usepackage{amssymb}
\usepackage{latexsym}
\usepackage[round]{natbib}
\usepackage{url}
\usepackage{amsmath}
\usepackage{bm}

\usepackage[final]{graphicx}
\usepackage[lofdepth,lotdepth]{subfig}

\usepackage{float}

\usepackage{xcolor}
\usepackage{float}
\restylefloat{table}

\usepackage{algorithm}
\usepackage[noend]{algpseudocode}

\usepackage[title]{appendix}

\newcommand{\prob}{\mathrm{Pr}}

\newcommand{\g}{\,\vert\,}
\newcommand{\bc}{\begin{center}}
\newcommand{\ec}{\end{center}}

\newcommand{\bitem}{\begin{itemize}}
\newcommand{\eitem}{\end{itemize}}

\newcommand{\be}{\begin{eqnarray*}}
\newcommand{\ee}{\end{eqnarray*}}
\newcommand{\ben}{\begin{eqnarray}}
\newcommand{\een}{\end{eqnarray}}

\usepackage{layout}

\renewcommand{\baselinestretch}{2}

\newcommand{\black}{\color{black}}

\begin{document}

\title{Objective Bayesian Comparison of  Order-Constrained Models in Contingency Tables }

\author{
Roberta Paroli\\
Universit\`{a} Cattolica del Sacro Cuore, Milan, Italy\\
\url{roberta.paroli@unicatt.it}
\and
Guido Consonni \\
Universit\`{a} Cattolica del Sacro Cuore, Milan, Italy\\
\url{guido.consonni@unicatt.it}
}
\date{  }    
\maketitle

\begin{center}
\end{center}

\begin{abstract}

In social and biomedical sciences testing in contingency tables often involves order restrictions on cell-probabilities parameters.
We develop objective Bayes methods for order-constrained testing 
and model comparison 
when observations arise under product binomial or multinomial sampling.
Specifically, we consider tests for monotone order of the parameters against equality of all parameters. Our strategy combines in a unified way both the intrinsic prior methodology and the encompassing prior approach in order to compute  Bayes factors and posterior model probabilities. Performance of our method is evaluated on several simulation studies and real datasets.

\end{abstract}


\section{Introduction}

 Taking into account the   ordering of categories in the analysis of two-way contingency tables may lead to improvements both in terms of power and model parsimony; see \citep{Agre:Coull:2002}.
Often ordered categories can  be naturally  associated with inequality constraints
among cell-probabilities,
 leading to substantial improvements over  models which ignore the ordinal information.

\citet{Agre:Coull:2002} provide
an extensive survey of the analysis of contingency tables under
inequality constraints from a frequentist perspective.

Over the years a growing  dissatisfaction has emerged among statisticians  over  conventional
measures of evidence such as $p$-values, as dramatically exemplified in \citet{John:etal:2017} with special reference to psychological studies.
In parallel Bayesian methods for hypotheses testing have become increasingly popular among practitioners; see again \cite{Wage:2007}  with reference to Psychology where the issue of replication studies is especially critical.

In particular the Bayes Factor
 (\cite{Kass:Raft:1995}) has emerged as a powerful tool for testing hypotheses (not necessarily nested) and model comparison. Additionally, if supplemented
with prior model probabilities it leads to a full posterior distribution on the set of models under consideration, which entirely summarizes  inference; see \cite{Ohag:Fors:2004}, ch 7.
This is a rich and informative output which provides an appreciation of the  strengths of the various models, as well as of the associated uncertainty.

Bayesian testing for contingency tables dates back to the works of Good and co-authors, e.g.
\cite{Good:1967, Crook:Good:1980, Good:Crook:1987}, and was mostly focused on  testing  independence against  an unrestricted  hypothesis. The latter problem was approached from an objective Bayes perspective in \citet{Case:More:2009}, while
more specialized settings were discussed
in \citet{Cons:LaRo:2008}; \citet{Ili:Kat:Ntz:2009}; \citet{Cons:More:Vent:2011}.
Testing of inequality-constrained hypotheses   were initially dealt with  in a series of papers with a  focus on psychology studies; see for instance the review article \citet{Hoij:2013}. Further analyses  were presented
 \citet{Bart:Scac:Farc:2012} and \citet{Kate:Agre:2013}.

All the previous papers relied on some form of subjectively specified (possibly weakly informative) priors.
Over the years however the objective Bayes method  has emerged as a powerful tool both for inference \citep{Berg:2006} and model choice
\citep{Pericchi:2005,  Cons:etal:2018} where the prior is determined by formal rules which are model-dependent but otherwise are free form subjective elicitation. This turns out to be especially advantageous in model comparison, where the influence of the prior distribution is notoriously pervasive and persistent even with increasing sample size.
This paper presents an objective   Bayes methodology  for the comparison of models for contingency tables specified by inequality constraints.
In particular we follow an intrinsic prior approach \citep{Berg:Peri:1996,More:1997, Pere:Berg:2002}, coupled  with an encompassing prior approach, as we detail in the paper.
%

Specifically, we consider two scenarios.
The first one concerns a collection of independent binary responses over  $r$-ordered levels of a factor/predictor, so
   that the underlying sampling model is product binomial.
  Interest centers on testing  the equality of the  probabilities of success against a monotone ordering. We focus on binary responses for simplicity of exposition, although our methods could be conceptually extended   to situations with polytomous responses (product multinomial model).
In the  second scenario we assume a joint multinomial model for the collection of cell frequencies, and test independence of rows and columns against inequality-constrained hypotheses on sets of cell probabilities, or functions thereof such as suitable odds-ratios.

%

The paper is organized as follows.
Section \ref{sec:Hyp test prod binom} presents the product binomial model focusing on the comparison between the null model of equal probabilities and the full model; it also discusses conventional and intrinsic priors  for this problem.
  Section \ref{sec:IEApp} is devoted to the comparison of constrained  product binomial models and contains the main contribution of the paper, named intrinsic-encompassing approach.  Section \ref{sec:Hyp test multin} implements our procedure  on   the multinomial model. Section \ref{sec:Applications} presents some simulations and real applications in medical and psychological studies. Finally Section \ref{sec:concluding} offers some points for discussion.

\section{Hypothesis testing in the product binomial model}
\label{sec:Hyp test prod binom}

\subsection{Notation and likelihood}
\label{subsec:notation}
Let $U$ be a binary response variable, and   $Z$ a factor  having $r$ ordered levels, $z_i$, $i=1,\ldots,r$.
Let $\pi_i=\prob\{U=1|Z=z_i, \pi_i\}$ be the probability of a success at level $Z=z_i$, $i=1,\ldots,r$.
If a random sample of  $n_i$ responses at level $z_i$ is available,
denote with
$Y_i$  the number of successes out of the $n_i$ trials. Then, conditionally on $\pi_i$, $Y_i$ is $\mbox{Binomial}(n_i, \pi_i)$.
If the $Y_i$'s are assumed to be independent, their sampling
distribution - and by extension that of the allied
$r \times 2$ contingency table containing
the frequencies $\{(Y_i, n_i-Y_i), \, i=1,\ldots,r \}$ -
is \textit{product Binomial}.

We now discuss briefly,
for completeness  and for later use the standard  setting, wherein
interest centers on  testing the null model (hypothesis) of \textit{equality} of success probabilities  across levels of $Z$
\ben
\label{eq:M0}
M_0: \pi_1=\pi_2=\ldots =\pi_r=\pi^*
 \een
against the   {\it encompassing}  model
\ben
\label{eq:Me}
M_e: (\pi_i, i=1,\ldots,r )\in \{ [0,1]^r\setminus \{\pi_1 =\ldots =\pi_r\} \}.
\een

Notice that $M_e$ imposes no restriction on the collection of probabilities $\{  \pi_i\}$ save for barring the possibility of  complete equality. For this reason $M_e$ could also be named \textit{unconstrained}; however we prefer the term encompassing for reasons that will become clear later on.

Let $\bm{y} = (y_i, i=1,\ldots,r)$, and set
$\bm{\pi}=(\pi_1,\ldots,\pi_r)$. The sampling distribution of $\bm{Y}$ for given $(n_{i}, i=1,\ldots,r)$ under $M_0$, respectively $M_e$, is
 \ben
 \label{samplingdistrib-0}
f(\bm{y}|\pi^*, M_0)= \prod_{i=1}^{r} \binom{n_i}{y_i} {\pi^*}^{y_i} (1-\pi^*)^{n_i-y_i} = K(\bm{n},\bm{y})\left\{ {\pi^*}^{s_y} (1-\pi^*)^{n-s_y}\right\}
\een
\ben
 \label{samplingdistrib-e}
f(\bm{y}|\bm{\pi}, M_e)= \prod_{i=1}^{r} \binom{n_i}{y_i} \pi_i^{y_i} (1-\pi_i)^{n_i-y_i} = K(\bm{n},\bm{y}) \left\{\prod_{i=1}^{r}\pi_i^{y_i} (1-\pi_i)^{n_i-y_i}\right\}
\een
where $K(\bm{n},\bm{y})=\prod_{i=1}^{r} \binom{n_i}{y_i}$, $n=\sum_{i=1}^{r}n_{i1}$,  $s_y=\sum_{i=1}^{r}y_i$ and $\bm{n}=(n_1,\ldots,n_r)$.

\subsection{Conventional priors}
\label{subsec:conventional priors}

Denote with
$p^N(\bm{\pi}|M_e)$
an objective prior for $\bm{\pi}$ under model $M_e$. Typically, this will be a reference \citep{Bern:1979}, or default,   prior  used for estimation purposes. The superscript \lq \lq $N$\rq \rq{} stands for
\textit{noninformative}.
A natural family for such a prior is a product of Beta distributions
\be
p^N(\bm{\pi}|M_e)= \prod_{i=1}^r Beta(\pi_i|\alpha_{i1}, \alpha_{i2}).
\ee
Now let $\pi^*$ be the success probability common to all levels under $M_0$.
A default prior is
\be
p^N(\pi^*|M_0)= Beta(\pi^*|\alpha_{01}, \alpha_{02}).
\ee
A standard choice might be
$\alpha_{i1}=\alpha_{i2}=1$, for $i=1,\ldots,r$, and  $\alpha_{01}=\alpha_{02}=1$, corresponding to a uniform prior; alternatively one could choose the value 1/2 as in the Jeffreys prior.

The  marginal likelihood under each of the two models is given by
\ben
\label{mN-0}
&& m^{N}(\bm{y}|M_0)
=\int_{0}^{1}
f(\bm{y}|\pi^*, M_0)p^{N}(\pi^*|M_0) d\pi^*
 \nonumber\\
&=&
K(\bm{n},\bm{y})\frac{B(\alpha_{01}+s_y, \alpha_{02}+n-s_y)}{B(\alpha_{01}, \alpha_{02})}
\een
and
\ben
\label{mN-e}
&& m^{N}(\bm{y}|M_e)
=\int_{0}^{1}
f(\bm{y}|\bm{\pi}, M_e)p^{N}(\bm{\pi}|M_e) d\bm{\pi}
\nonumber\\
&=&
K(\bm{n},\bm{y}) \prod_{i=1}^{r} \frac{B(\alpha_{i1}+y_i, \alpha_{i2}+n_i-y_i)}{B(\alpha_{i1}, \alpha_{i2})}.
\een
They are subsequently employed to produce the \textit{Bayes factor} (BF) of $M_e$ against $M_0$ which is given by
\ben
\label{BF^N}
BF_{e0}^N(\bm{y})=\frac{m^{N}(\bm{y}|M_e)}{m^{N}(\bm{y}|M_0)},
\een
where the superscript $N$ is used to remind us that the BF is computed using the default priors, and to distinguish it from an alternative BF we shall employ later on.

\subsection{Intrinsic priors}
\label{subsec:intrinsic priors}

It is by now an established fact within the Bayesian community  that \textit{objective} priors, which have been designed for estimation purposes \textit{conditionally on a given model},  are largely inadequate for model comparison or hypotheses testing; see  \citet{Pericchi:2005} and \citet{Cons:etal:2018}. 
This is patently evident when the objective prior under any of the two models is improper, because the presence of an arbitrary normalizing constant in the prior  transfers to the marginal likelihood and consequently makes the $BF^N$ meaningless.
However the rationale for \textit{not} using conventional objective priors for testing holds also when the prior under each of the two models is \textit{proper}, as in our case. The reason for this is that $p^N(\bm{\pi} |M_e )$ is not \textit{compatible} with $p^N(\pi |M_0 )$, i.e.  it is  not chosen in view of
the comparison with model $M_0$. For more information  on the issue of compatibility  of priors for model selection see \citet{Cons:Vero:2008}.


  In particular, a conventional objective prior $p^N(\bm{\pi} |M_e )$  is generally diffuse, and  thus gives relatively little weight to parameter values close to the subspace characterizing $M_0$. Consequently,
there is an evidence bias in favor of $M_0$ (unless the data are vastly against $M_0$, which rarely happens for moderate
sample sizes).  Informally, one can say that $p^N(\cdot |M_e )$ \lq \lq wastes\rq \rq{} away probability mass in parameter areas too remote
from the null.
To overcome this difficulty, one ought to modify $p^N(\cdot |M_e )$ so that it reallocates more probability mass
toward the null subspace, an idea already advocated in \citet[Chapter 3]{Jeff:1961}.
 This of course has a negative side effect, at least for moderate sample sizes, because  it will diminish evidence
in favor of $M_e$ when the parameter values generating the data are truly away from the null. However this is a price worth paying, as
 explicated in \citet{cons:etal:2013}, to whom we refer the reader for further considerations about issues discussed in this subsection.

We now describe a strategy to implement the above program based on the notion of  \textit{intrinsic priors},
which were introduced in objective hypothesis testing  to deal in a sensible way with \textit{improper} default
priors; see \citet{Berg:Peri:1996} and
\citet{More:1997}.
 However the  scope of the intrinsic prior approach is much wider, because it represents a general methodology for Bayesian model choice, and can be used in any circumstance, and so also when the  starting default priors are proper, as in our case.
 The reason why intrinsic priors are especially effective is easily seen
when comparing two nested models, such as $M_0$ and  $M_e$ in Section \ref{sec:Hyp test prod binom}.
The basic idea is to introduce a set of \textit{imaginary} observations, i.e. auxiliary random variables,  to  \lq \lq train\rq \rq{} the default prior  $p^N(\cdot |M_e )$ so that its diffuseness is reduced by
shifting some probability mass toward the null subspace characterizing $M_0$.
Let us see how we can achieve this goal in the setup of a product binomial model.
Let  $\bm{x}=(x_i, i=1,\ldots,r)$ be imaginary observations,  with $x_i$ representing the number of  successes out of $t_i$ trials, and set
 $\bm{t}=(t_1,\ldots,t_r)$.
The intrinsic prior under model $M_e$ for the comparison with model $M_0$ is defined as
\ben
\label{I}
p^I(\bm{\pi}|\bm{t},M_e)= \sum_{\bm{x}} p^N(\bm{\pi}|\bm{t},\bm{x},M_e) m^N(\bm{x}|\bm{t}, M_0),
\een
where
$p^N(\bm{\pi}|\bm{t},\bm{x},M_e)\propto p^N(\bm{\pi}|M_e) f(\bm{x}|\bm{t}, \bm{\pi},M_e)$, so that
\ben
\label{eq:p^N(pi|t,x,Me)}
p^N(\bm{\pi}|\bm{t},\bm{x},M_e)
=
\prod_{i=1}^r Beta(\pi_i|\alpha_{i1}+x_i;\alpha_{i2}+t_i-x_i).
\een
On the other hand
\be
m^N(\bm{x}|\bm{t},M_0)=K(\bm{t},\bm{x}) \frac{B(\alpha_{01}+s_x, \alpha_{02}+t-s_x)}{B(\alpha_{01}, \alpha_{02})}
\ee
is the marginal distribution of $\bm{X}$ under model $M_0$ with prior $p^N(\cdot |M_0)$ which can be seen to be the analogue of (\ref{mN-0}) upon replacing $\bm{y}$ with $\bm{x}$ and $\bm{n}$ with $\bm{t}$, so that  $s_x=\sum_{i=1}^{r}x_i$ and  $t=\sum_{i=1}^{r} t_i$.

\begin{figure}
\subfloat[$\bar{t}=3$]{\includegraphics[width = 3in]{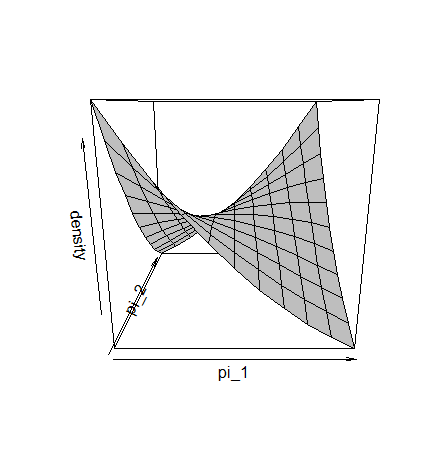}}
\subfloat[$\bar{t}= 5$]{\includegraphics[width = 3in]{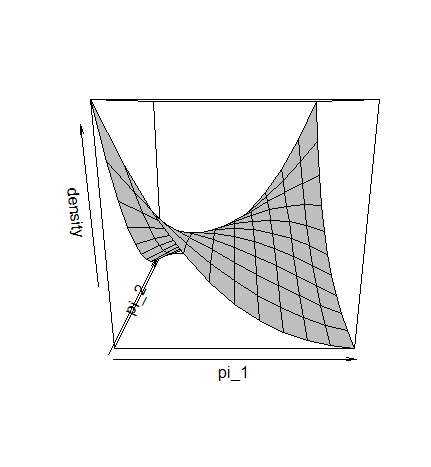}}\\
\subfloat[$\bar{t}=10$]{\includegraphics[width = 3in]{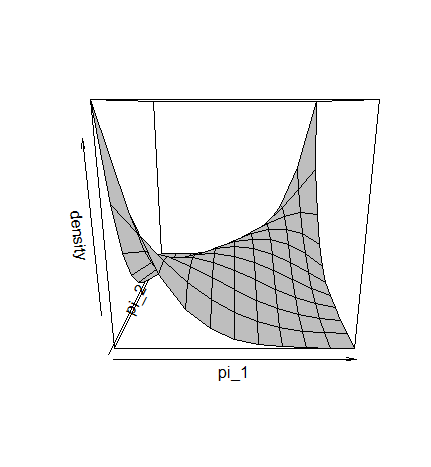}}
\subfloat[$\bar{t}=50$]{\includegraphics[width = 3in]{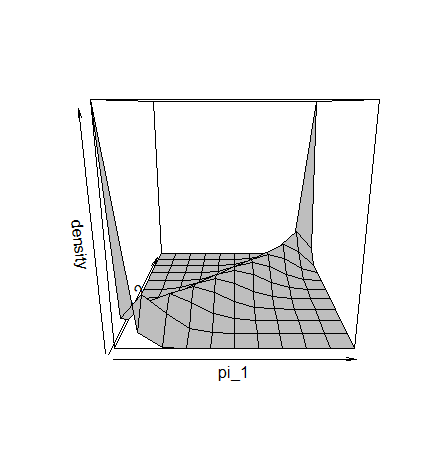}}
\caption{Intrinsic prior  for different values of the training sample size $\bar{t}$.}
\label{intr-prior}
\end{figure}

\textit{Remarks}
\begin{itemize}
\item
The intrinsic prior is a mixture of \lq \lq pseudo-posteriors\rq \rq{} $p^N(\bm{\pi}|\bm{x},M_e)$ with respect to the mixing distribution
$m^N(\bm{x}|M_0)$. As a consequence, the individual $\pi_i$'s, which were \textit{independent} under $p^N(\cdot |M_e)$ are no longer so under $p^I(\cdot |M_e)$.

\item
It can be checked that if the training sample size for row $i$,  $t_i$ is zero,   i.e. if no intrinsic procedure is applied, then marginally $p^I(\pi_i \g t_i=0, M_e)=p^N(\pi_i)$, i.e. the intrinsic prior reduces to the initial prior which is recovered as a special case.
On the other hand, as each of the $t_i$  increases,
 $p^I(\bm{\pi}|\bm{t}, M_e)$ will  transfer more mass  to the one-dimensional subspace $\pi_1=\pi_2=\dots, \pi_r$, so that in the limit $p^I(\bm{\pi}|\bm{t},  M_e)$ will degenerate to a uniform distribution on that subspace:
  see Fig. \ref{intr-prior} for an illustration of this phenomenon.
  As a consequence  the intrinsic marginal data distribution  $m^I(\bm{y} |\bm{t}, M_e)=\int f(\bm{y} |\bm{\pi}, M_e) p^I(\bm{\pi} |\bm{t}, M_e)$ will tend to $m^N(\bm{y}|M_0)$, and the corresponding BF will converge to one.
 One can thus see that the choice of the $t_i$'s is quite important in comparing the two models because it
 regulates the
amount of concentration of the intrinsic prior around the null subspace.
To circumvent this difficulty, it is customary to set $t_i$ equal to a \textit{minimal  training sample size} \citep{Berg:Peri:AS:2004} which guarantees that the posterior is proper. In our setup however, because our starting default priors are already {proper} under each of the two models,  the notion of minimal training sample size becomes  useless. Accordingly,  we will let each $t_i$ range over the integers in the set $\{ 0, 1, \ldots,n_i \}$, thus effectively performing a sensitivity analysis. This means that if the results do not change appreciably as $t_i$ varies,  then our inferential conclusion is robust.

%
%

%

\item
The intrinsic prior in (\ref{I}) is a special case of the \textit{expected posterior prior} introduced in \citet{Pere:Berg:2002} for the comparison of several models $M_k$ each equipped with a default prior $p^N(\cdot |M_k)$. In that case  the intrinsic prior under $M_k$ is as in (\ref{I}) with the mixing distribution $m^N(\cdot | M_0)$  replaced by a more general measure $m^*(\cdot)$, which however must be the same for all models.

\item
 A very simple expression of the intrinsic prior can be written under the constrained hypothesis, due to the exchangeability property, when the sizes of the training samples are equal $t_1=t_2=\ldots=t_r$ (see the Supplementary Material).
\end{itemize}

An illustration of  the behavior of the intrinsic prior in the simple case of $r=2$ is provided in Figure   \ref{intr-prior} for different training sample sizes $\bar{t}=t_1=t_2 $, and with all hyperparameters $\alpha$'s  set to 1. Notice that as $\bar{t}$ increases, the intrinsic prior progressively  concentrates around the line $\pi_1=\pi_2$.

\section{Comparison of constrained product binomial models }
\label{sec:IEApp}

 \citet[Table 1]{Agre:Coull:2002} discuss a clinical trial applied to  patients who experienced trauma due to subarachnoid hemorrhage.
  Factor $Z$ has four levels, corresponding to a placebo followed by  three increasing doses of a medication. The outcome variable
 has five levels
 (\lq \lq Death\rq \rq{}, \lq \lq Vegetative state\rq \rq{}, \lq \lq Major disability\rq \rq{},\lq \lq Minor disability \rq \rq{},\lq \lq Good recovery\rq \rq{})
  but for illustration purposes it has been collapsed
 to a binary variable, with categories \lq \lq Death\rq \rq{} and  \lq \lq Not Death\rq \rq{}.
 The resulting $4 \times 2$ contingency table is reported in Table \ref{tab:tabA&C}.

{\renewcommand{\baselinestretch}{1}
\begin{table}[!ht]
\centering \caption{\protect\small \textit{Responses from a clinical trial comparing four treatments on patients who experienced trauma due to subarachnoid hemorrhage}}
 \label{tab:tabA&C}
\begin{tabular}{|c|c|c|}
\hline
 Treatment & \multicolumn{2}{|c|}{Outcome} \\
 & Death & Not Death \\
\hline
Placebo     & 59 & 151\\
Low dose    & 48 & 142\\
Medium dose & 44 & 163 \\
High dose   & 43 & 152\\
\hline
\end{tabular}
\end{table}
}

It is expected that a more favorable outcome tends to occur as the dose increases.  Taking this information into account,
the null model $M_0$  in (\ref{eq:M0})  is tested against the model based on an ordered restriction of the probabilities of Death $\pi_i$:
\ben
\label{eq:Mc}
M_c: \pi_1> \pi_2> \ldots > \pi_r.
\een

In the sequel $M_c$ will denote a generic \textit{constrained} model containing inequalities constraints such as those in (\ref{eq:Mc}). More general types of constrained models could be envisaged, such as those containing a mixture of equality and inequality constraints; see for instance \citet{MULDER:2014}.

To perform this comparison using the Bayes factor we require a prior on the parameter space under model $M_c$. This can be achieved using an encompassing prior approach \citep{Klug:Hoij:2007}, which we now briefly summarize.
Let $M_c$ be a constrained model whose parameter space $\bm{\Theta_c}$  is specified by means of  inequalities on the components of $\bm{\theta} \in \bm{\Theta} $, with
$\bm{\Theta} $ being the  unconstrained parameter space of the   encompassing model $M_e$.
Let $p(\bm{\theta}|M_e)$ be a (proper) prior under
$M_e$. A natural way to construct a prior under $M_c$ is by truncation, namely $p(\bm{\theta}|M_c) \propto p(\bm{\theta}|M_e) \bm{1}_{\bm{\Theta}_c}(\bm{\theta})$, where $\bm{1}_{A}(\cdot)$ is the indicator function of the set $A$.
A straightforward calculation \citep{Klug:Hoij:2007} shows that,  for fixed data $\bm{y}$, the Bayes factor of $M_c $ against $M_e$, is given by the ratio
${\prob \{ \bm{\theta} \in \bm{\Theta}_c |\bm{y}, M_e \}}/{\prob \{ \bm{\theta} \in \bm{\Theta}_c | M_e \}}$.
Note that both  the prior and the  posterior probability of the set $\bm{\Theta}_c$ are evaluated under the unconstrained model $M_e$. \citet{WetzelsEtAl:2010}
provide further  comments on the encompassing prior approach.

Having specified the theoretical framework we work with, our strategy to construct an objective Bayes factor of model $M_c$, specified by the constrained parameter space  $\bm{\Theta}_c$,  as for instance in (\ref{eq:Mc}),   against $M_0$ can be  outlined as follows:
\begin{itemize}
\item
start with an objective   prior $p^N(\bm{\pi} |M_e)$, which we assume to be proper;
\item
for given training sample sizes $\bm{t}$, construct the intrinsic prior $p^I(\bm{\pi} |\bm{t}, M_e)$ tailored to the comparison of $M_e$ against $M_0$ as in (\ref{I}),  and derive the BF based on the intrinsic prior (which we label as $BF^I$)
\ben
\label{eq:BF^Ie0}
BF^{I}_{e0}(\bm{y} | \bm{t})=\frac{m^{I}(\bm{y}|\bm{t}, M_e)}{m^N(\bm{y}|M_0)},
\een
where $m^{I}(\bm{y}|\bm{t}, M_e)=\int f(\bm{y} |\bm{\pi}, M_e) p^I(\bm{\pi} |\bm{t}, M_e) d\bm{\pi}$;
\item
compute
\ben
\label{eq:BF^Ice}
BF^{I}_{ce}(\bm{y} | \bm{t})=
\frac{\prob^{I} \{ \bm{\pi} \in \bm{\Theta}_c | \bm{t}, \bm{y}, M_e \}}{\prob^{I} \{ \bm{\pi} \in \bm{\Theta}_c |\bm{t}, M_e \}};
\een

\item
finally derive
\ben
\label{eq:BF^Ic0}
BF^{I}_{c0}(\bm{y} | \bm{t})=BF^{I}_{ce}(\bm{y} | \bm{t})
\times BF^{I}_{e0}(\bm{y} | \bm{t}).
\een
\end{itemize}
The above procedure, which we name \textit{intrinsic-encompassing}, was first presented in
\citet{Cons:Paro:2017} with regard to the comparison of constrained ANOVA models.
There is however a significant difference.
In that setting  the starting default priors were improper, so that in particular the intrinsic prior under $M_e$ was also improper; accordingly the procedure had to be based on the \textit{conditional} intrinsic prior (which is always proper), rather than the actual  intrinsic prior, as  in our current setup. 
The main advantage of the
{intrinsic-encompassing} approach is contained in formula (\ref{eq:BF^Ic0}). It can be seen that the computation of
$BF^{I}_{c0}(\bm{y} | \bm{t})$ is decoupled
into two parts: i) one involving the computation of the \lq \lq standard\rq \rq{} Bayes factor $BF^{I}_{e0}(\bm{y} | \bm{t})$, and  ii) one involving the evaluation of two probabilities of the set $\bm{\Theta}_c$. While the latter formally requires  integrating over $\bm{\Theta}_c$, a simulation-based approximation
is typically available  based on  draws from the intrinsic prior, respectively posterior,  under the unconstrained model $M_e$.

%
%
%

Assuming that the true model belongs
to a finite  space of (not necessarily nested) constrained models $\{ M_c: c \in {\cal C} \}$, the above procedure can be used to obtain the posterior distribution on model space,  provided one can identify a single null model $M_0$ which is nested into any model under consideration.
In that case one gets
\ben
\label{eq:prob^IM_c}
\prob^I(M_c | \bm{t}, \bm{y})=
\frac{BF^{I}_{c0}(\bm{y}|\bm{t})}{1+(\sum_{c^{\prime} \in {\cal C}} BF^{I}_{c^{\prime}0}(\bm{y}|\bm{t}) P_{c^{\prime}0}    ) }, \, c \in {\cal C},
\een
where $P_{c0}=\prob(M_c)/\prob(M_0)$ is the prior odds  of model $M_c$ against model $M_0$.
Notice that the calculation leading to (\ref{eq:prob^IM_c}) is coherent because
 the marginal distribution of the data under $M_0$, $m^N(\bm{y} |M_0)$, is
the same under any  Bayes factor involved in (\ref{eq:prob^IM_c}).
This means in particular that the BF for the comparison
of models $M_c$ and $M_{c^{\prime}}$   is computable as  $BF^{I}_{c c^{\prime}}(\bm{y}|\bm{t})=BF^{I}_{c0}(\bm{y}|\bm{t})/BF^{I}_{c^{\prime}0}(\bm{y}|\bm{t})$.
It is important to realize that the posterior probability of model $M_c$ will depend  not only on the fit to the observations but also to the model complexity. This is because the Bayes factor incorporates an automatic Ockham's razor
\citep{Jeff:Berg:1992},  whereby more complex models are implicitly discounted. Interestingly, this penalization applies not only to the standard comparison of two models having  different dimensionality (number of parameters) but also to models having the same dimension wherein one has a smaller parameter space. In particular our approach allows to meaningfully compare  an inequality constrained model with an unconstrained model, so that the former may receive a higher posterior probability than the latter as some examples below will clarify.

\subsection{BF of the encompassing model against the null model}

In this subsection we detail  calculations to obtain (\ref{eq:BF^Ie0}) in our setting.

The summations involved may cause computational problems when the number of groups $r$ and the dimensions of the training sample $t_i$  are large because the number of their terms becomes prohibitively large.
This difficulty can  be effectively overcome by means of Monte Carlo sum as described in \citet{Case:More:2005} and \citet{Cons:More:Vent:2011}.

On the other hand \citep[(4.2)]{Pere:Berg:2002} showed  that $BF^{I}_{e0}(\bm{y}|\bm{t})$ can be approximated using importance sampling as
\ben
 \label{BFe0Estimate}
\widehat{BF}^{I}_{e0}(\bm{y}|\bm{t})\simeq \frac{1}{S} \sum_{s=1}^{S}
\frac{m^N(\bm{y}| \bm{t},\bm{x}^{(s)},M_e)}{m^N(\bm{y}| \bm{t},\bm{x}^{(s)}, M_0)},
\een
where $\bm{x}^{(s)}$, $s=1, \ldots, S$,  are draws from the importance distribution $m^N(\bm{x}|\bm{t}, \bm{y},M_0)$.
The analytical form of the terms appearing in \eqref{BFe0Estimate} are specified  below.

\be
m^N(\bm{y}|\bm{t}, \bm{x},M_e)= K(\bm{n},\bm{y}) \prod_{i=1}^{r} \frac{B(\alpha_{i1}+x_i+y_i;\alpha_{i2}+(t_i-x_i)+(n_i-y_i))}{B(\alpha_{i1}+x_i;\alpha_{i2}+(t_i-x_i))},
\ee
\be
m^N(\bm{y}|\bm{t},\bm{x},M_0)= K(\bm{n},\bm{y}) \frac{B(\alpha_{01}+s_x+s_y;\alpha_{02}+(t-s_x)+(n-s_y))}{B(\alpha_{01}+s_x;\alpha_{02}+(t-s_x))}.
\ee
Finally the importance distribution is given by
\ben
\label{eq:importance-distribution}
m^N(\bm{x}|\bm{t}, \bm{y},M_0)= K(\bm{t},\bm{x}) \frac{B(\alpha_{01}+s_x+s_y;\alpha_{02}+(t-s_x)+(n-s_y))}{B(\alpha_{01}+s_y;\alpha_{02}+(n-s_y))}.
\een
To sample from \eqref{eq:importance-distribution} one can proceed as follows:

\bitem
\item[i)]
sample $\pi^{(s)}$  from the posterior distribution $p^N(\bm{\pi} |\bm{y}, M_0)$
\be
\pi^{(s)}|\bm{y},M_0\sim Beta(\alpha_{01}+s_y;\alpha_{02}+(n-s_y)),
\ee

\item[ii)]
sample each element of $\bm{x}^{(s)}$ independently
\be
x_i^{(s)}|t_i, \pi^{(s)},M_0 \sim Bin(x_i|t_i,\pi^{(s)}), \, i=1,\ldots,r.
\ee
\eitem

\subsection{BF of the constrained model against the  encompassing model }
\label{sec:BFce}
The expression for $BF^I_{ce}(\bm{y} | \bm{t})$ in (\ref{eq:BF^Ice}) is a ratio of probabilities for the same subspace $\bm{\Theta}_c$.
The denominator  involves the intrinsic prior, which  is is given by
\ben
\label{priorBFce}
p^I(\bm{\pi}|\bm{t},M_e)= \sum_{\bm{x}}
\left \{
\left[\prod_{i=1}^r Beta(\pi_i|\alpha_{i1}+x_i;\alpha_{i2}+t_i-x_i )\right]
 m^N(\bm{x}|\bm{t}, H_0)
\right \}.
\een
On the other hand,  the numerator 
involves the intrinsic posterior distribution, whose density can be written as
 \ben
\label{eq:intrinsic posterior}
& & p^I(\bm{\pi}|\bm{y},\bm{t}, M_e) \nonumber  \\
&=&
 \sum_{\bm{x}} \left\{\prod_{i=1}^r Beta(\pi_i|\alpha_{i1}+x_i+y_i;\alpha_{i2}+(t_i-x_i)+(n_i-y_i)) \right\}
 m^*(\bm{x} |\bm{t},\bm{y}),
 \een
 where
\be
 m^*(\bm{x} |\bm{t}, \bm{y})=\frac{m^N(\bm{y}|\bm{x}, \bm{t}, M_e)m^N(\bm{x}| \bm{t}, M_0)}{\sum_{\bm{x}}m^N(\bm{y}|\bm{x}, \bm{t}, M_e)m^N(\bm{x}| \bm{t}, M_0)}.
 \ee
We can write more compactly
\be
 m^*(\bm{x} |\bm{t},\bm{y})=\frac{H(\bm{x} |\bm{t},\bm{y})}{\sum_{\bm{x}}H(\bm{x} |\bm{t},\bm{y})},
 \ee
where
\ben
\label{eq:H(x|t,y)}
 H(\bm{x} |\bm{t},\bm{y}) &=&
  K(\bm{t},\bm{x}) \left[ \prod_{i=1}^r \frac{B(\alpha_{i1}+x_i+y_i;\alpha_{i2}+(t_i-x_i)+(n_i-y_i))}{B(\alpha_{i1}+x_i,\alpha_{i2}+(t_i-x_i))} \right] \nonumber \\
 &\times&    B(\alpha_{01}+s_x, \alpha_{02}+t-s_x),
\een
since the function $H(\bm{x} |\bm{t},\bm{y})$ will be used later on in our computations.




Both the intrinsic prior and posterior  are discrete mixtures of product of Beta distributions with respect to the imaginary observations $\bm{x}$. In both cases the number of terms in the sum can be prohibitively large; additionally, for each fixed $\bm{x}$, integration over $\Theta_c$ is typically not analytically available.
To address the above difficulties, we can use an importance sampling strategy, as for instance implemented
 in \citet{Case:More:2009} and \citet{Cons:More:Vent:2011} in the context of intrinsic priors involving discrete mixtures.

Consider first the evaluation of the \textit{denominator} of $BF^I_{ce}$. We approximate the required probability by drawing $S$ independent and identically distributed  samples from the intrinsic prior \eqref{priorBFce}. The latter in turn can be regarded as the marginal distribution of
$\bm{\pi}$, derived from the joint distribution
\be
p(\bm{\pi}, \pi, \bm{x}  )=p^N(\pi |M_0)f(\bm{x} |\pi,\bm{t}, M_0) p^N(\bm{\pi}|\bm{t}, \bm{x}, M_e).
\ee
 Each of these three components can be sampled iteratively, and in the end we retain only the  $\bm{\pi}$-values (see Algorithm \ref{alg:den12}).

Consider now the \textit{numerator} of $BF^I_{ce}$.
Since it is not possible to sample exactly from the intrinsic posterior distribution, we rely on
a Metropolis within Gibbs algorithm as in Algorithm \ref{alg:num12}.

Finally obtain
\be
\label{BFceEstimate}
\widehat{BF}^{I}_{c,e}(\bm{y}|\bm{t})=
\frac{\widehat{\prob}^{I} \{\pi_1>\pi_2> \ldots > \pi_r|\bm{y},\bm{t},M_e  \}}
{\widehat{\prob}^{I} \{ \pi_1>\pi_2> \ldots > \pi_r|\bm{t}, M_e  \}}.
\ee

\begin{algorithm}
	\caption{Approximation of the denominator of \eqref{eq:BF^Ice}  }\label{alg:den12}
	\begin{algorithmic}[1]
		\For{$s=1,2,\ldots,S$}
		\State sample $\pi^{(s)}$ from $p^N(\pi |M_0)$:
		$\pi^{(s)}\sim Beta(\alpha_{01},\alpha_{02})$;
		\State sample independently  each element $x_i$ of $\bm{x}^{(s)}$ from $f({x_i} |\pi^{(s)},{t_i}, M_0)$:	$x_i^{(s)}|\pi^{(s)}, t_i,M_0 \sim Bin(x_i|t_i,\pi^{(s)})$;
		\State 	sample independently each element
			${{\pi_i}^{(s)}}$
			of $\bm{\pi}^{(s)}$
			from the pseudo posterior
			$p^N({\pi_i}|{t_i}, {x_i^{(s)}}, M_e)$:
			$\pi_i^{(s)}|{x_i}^{(s)},M_e \sim Beta(\pi_i |\alpha_{i1}+x_i^{(s)}, \alpha_{i2}+t_i-x_i^{(s)})$;
		\EndFor
		\State \textbf{end for}
		\State Approximate the probability as
		\be \widehat{\prob}^{I} \{ \pi_1>\pi_2> \ldots > \pi_r |M_e \} \thickapprox\frac{1}{S}\sum_{s=1}^{S} \bm{1}_{\pi_1^{(s)}>\pi_2^{(s)}> \ldots > \pi_r^{(s)}}(\bm{\pi}^{(s)}).\ee
		\end{algorithmic}
\end{algorithm}


\begin{algorithm}
	\caption{Approximation of the numerator of \eqref{eq:BF^Ice}  }\label{alg:num12}
	\begin{algorithmic}[1]
		\State estimate the probabilities $\pi_i$, $i=1,2,\ldots,r$  as
		\be \widehat{\pi_i}=\frac{y_i+1}{s_y+r}, \forall i=1,2,\ldots,r;\ee
		\State generate the imaginary training sample elements with this step of Metropolis:
		
		\State sample the starting values
		$x_i^{(0)}\sim Bin(t_i,\widehat{\pi_i})$, $\forall i=1,\ldots,r$,
		independently;
		\For{$s=1,2,\ldots,S_1$}
		\State generate  $\bm{x}^{(s)}$ by sampling
		each component $x_i^{(s)}$ from the proposal $Bin(t_i,\widehat{\pi_i})$,
		independently for $ i=1,\ldots,r$,
		and accept it with probability:
		\be
		\alpha\left(\bm{x}^{(s-1)};\bm{x}^{(s)}\right) =\min \left\{ 1;%
		\frac{H(\bm{x}^{(s)}| \bm{t},\bm{y}) \prod_{i=1}^{r} Bin({x}_{i1}^{(s-1)}|t_i,\widehat{\pi_i})
		}{H(\bm{x}^{(s-1)}| \bm{t},\bm{y})
			\prod_{i=1}^{r} Bin({x}_{i1}^{(s)}|t_i,\widehat{\pi_i}) }
		\right\};
		\ee
		where $H(\bm{x}| \bm{t},\bm{y})$ is defined in \eqref{eq:H(x|t,y)}.
		\EndFor
		\State \textbf{end for}
		\State 	after a suitable number of  burn-in iterations $S_1$, start sampling
		$\bm{\pi}^{(s)}$:
		\For{$s=S_1+1,\ldots,S_1+S$}
		\State repeat step (5) with the following addition:
		sample each component $\pi_i^{(s)}$ of  $\bm{\pi}^{(s)}$ from
		$
		\pi_i^{(s)}|t_i, {x}_{i1}^{(s)},M_e \sim Beta(\alpha_{i1}+x^{(s)}_{i1},\alpha_{i2}+t_i-x^{(s)}_{i1}),
		$
		independently for $i=1, \ldots, r$;
		\EndFor
		\State \textbf{end for}
		\State approximate the probability of the numerator of \eqref{eq:BF^Ice} as
		\be \widehat{\prob}^{I} \{ \pi_1>\pi_2> \ldots > \pi_r |\bm{y},\bm{t},M_e \} \thickapprox\frac{1}{S}\sum_{s=S_1+1}^{S_1+S} \bm{1}_{\pi_1^{(s)}>\pi_2^{(s)}> \ldots > \pi_r^{(s)}}(\bm{\pi}^{(s)}).
		\ee
	\end{algorithmic}
\end{algorithm}

\newpage


\section{Hypothesis testing in the multinomial model}
\label{sec:Hyp test multin}
In this section we extend the scope of  our methodology  to  an $r \times c$ contingency table under a multinomial sampling model.
 Denote with  $\bm{y}=\{y_{ij}, i=1,\ldots,r, j=1,\ldots,c \}$, with $\sum_{i=1}^r\sum_{j=1}^c y_{ij}=n$ the cell frequencies.
Under the encompassing model $M_e$ all cell probabilities are unconstrained, save for adding up to one, and we write  $\bm{\pi}=\{\pi_{ij}, i=1,\ldots,r, j=1,\ldots,c \}$ and $\sum_{i=1}^r\sum_{j=1}^c \pi_{ij}=1$.
A default prior for $\bm{\pi}$ is Dirichlet with hyperparameters $\bm{\alpha}_\pi$
$
p^N(\bm{\pi}|M_e)= Di(\bm{\pi}|\bm{\alpha}_\pi).
$
Typically $\bm{\alpha}_\pi$ has all its $rc$ elements equal to 1 (Uniform prior) or equal to $1/2$ (Jeffreys prior).
For given $n$, the sampling distribution of $\bm{Y}$  under  model $M_e$ is
 \ben
 \label{Msamplingdistrib-e}
f(\bm{y}|\bm{\pi}, M_e)=  \binom{n}{\bm{y}} \prod_{i=1}^{r} \prod_{j=1}^{c} \pi_{ij}^{y_{ij}},
\een
where $\binom{n}{\bm{y}}$ is the multinomial coefficient.

Let $\bm{\pi_R}$ and $\bm{\pi_C}$ be the vectors of row, respectively column,  marginal probabilities, with $\sum_{i=1}^r \pi_{Ri}=1$ and $\sum_{j=1}^c \pi_{Cj}=1$. Denote the null model  of independence by
\ben
\label{eq:MM0}
M_0: \pi^0_{i,j}=\pi_{Ri} \cdot \pi_{Cj}, \quad i=1,\ldots,r; \,  j=1,\ldots,c
 \een
A default prior on $(\bm{\pi_R}, \bm{\pi_C})$ is
\be
p^N(\bm{\pi_R}, \bm{\pi_C}|M_0)= Di(\bm{\pi_R}|\bm{\alpha}_{\pi_R})\cdot Di(\bm{\pi_C}|\bm{\alpha}_{\pi_C}),
\ee
Typically both $\bm{\alpha}_{\pi_R}$ and $\bm{\alpha}_{\pi_C}$ have each all elements equal to 1 (Uniform prior) or equal to $1/2$ (Jeffreys prior).
The likelihood function under $M_0$ is
\ben
\label{Msamplingdistrib-0}
f(\bm{y}|\bm{\pi_R},\bm{\pi_C}, M_0)= \binom{n}{\bm{y}} \left( \prod_{i=1}^{r} \pi_{Ri}^{y_{i+}}\right) \left(\prod_{j=1}^{c} \pi_{Cj}^{y_{+j}}\right),
\een
where $y_{i+}=\sum_{j=1}^{c} y_{ij}$ and $y_{+j}=\sum_{i=1}^{r} y_{ij}$.

The marginal likelihood for the null model is given by
\ben
\label{MmN-0}
m^{N}(\bm{y}|M_0)=
\binom{n}{\bm{y}}\frac{B(\bm{\alpha}_{\pi_R}+\bm{y_{R+}}) B(\bm{\alpha}_{\pi_C}+\bm{y_{+C}})}{B(\bm{\alpha}_{\pi_R})B(\bm{\alpha}_{{\pi_C}})}
\een
with $\bm{y_{R+}}=(y_{i+}, i=1,\ldots,r)$ and $\bm{y_{+C}}=(y_{+j}, j=1,\ldots,c)$, and where  $B$ stands for the \textit{multivariate Beta function}.

In the multinomial model the constrained model $M_c$ is often stated in terms of inequality constraints between sets of cell probabilities or functions of cell probabilities like  odds ratios; for more details  see  \citet{Agre:Coull:2002}.

\subsection{Intrinsic priors}
Let  $\bm{x}=(x_{ij}, i=1,\ldots,r, j=1,\ldots,c)$ be a matrix of imaginary observations,  with  $\sum_{i=1}^{r} \sum_{j=1}^{c}x_{ij} =t$.
The intrinsic prior under model $M_e$ for the comparison with model $M_0$ is given by
\be
p^I(\bm{\pi}|t,M_e)= \sum_{\{\bm{x}:\sum_{ij}x_{ij}=t\}} p^N(\bm{\pi}|t,\bm{x},M_e) m^N(\bm{x}|t, M_0),
\ee
where the "pseudo posterior"  is Dirichlet
\be
p^N(\bm{\pi}|t,\bm{x},M_e)= Di(\bm{\pi}|\bm{\alpha}_\pi+\bm{x}),
\ee
while $m^N(\bm{x}|t,M_0)$ is the marginal distribution of $\bm{X}$ under model $M_0$ with prior $p^N(\cdot |M_0)$ which can be seen to be the analogue of (\ref{MmN-0}) upon replacing $\bm{y}$ with $\bm{x}$ and $n$ with $t$.
The explicit expression for the intrinsic prior is
\ben
\label{MIp}
p^I(\bm{\pi}|t,\bm{x},M_e)= \frac{\Gamma(t+rc)\Gamma(r)\Gamma(c)}{\Gamma(t+r)\Gamma(t+c)} \sum_{\{\bm{x}:\sum_{ij}x_{ij}=t\}}
\binom{t}{\bm{x}} \frac{(\prod_{i}x_{i+}!)(\prod_{j}x_{+j}!)}{\prod_{i,j}x_{ij}!} \prod_{i,j}\pi_{ij}^{x_{ij}}
\een
with $x_{i+}=\sum_{j=1}^{c} x_{ij}$ and $x_{+j}=\sum_{i=1}^{r} x_{ij}$.

The marginal likelihood for model $M_e$ under the intrinsic prior is given by:
\ben
\label{eq:mult-m^IMe}
& & m^{I}(\bm{y}|t, M_e)=\int f(\bm{y} |\bm{\pi}, M_e) p^I(\bm{\pi} |t, M_e) d\bm{\pi}  \\
&=&
\frac{\binom{n}{\bm{y}}}{B(\bm{\alpha}_{\pi_1})B(\bm{\alpha}_{\pi_2})} \sum_{\{\bm{x}:\sum_{ij}x_{ij}=t\}}\binom{t}{\bm{x}}
B(\bm{\alpha}_{\pi_1}+\bm{x_{R+}}) B(\bm{\alpha}_{\pi_2}+\bm{x_{+C}}) B(\bm{\alpha}_{\pi}+\bm{x+y}) \nonumber
\een
where $\bm{x_{R+}}=(x_{i+}, i=1,\ldots,r)$ and $\bm{x_{+C}}=(x_{+j}, j=1,\ldots,c)$.

\subsection{Bayes Factor}
As described in Section \ref{sec:Hyp test prod binom},
to apply the intrinsic-encompassing procedure we need the expressions of  two Bayes Factors, namely $BF^I_{e0}(\bm{y}|t)$ and  $BF^I_{ce}(\bm{y}|t)$.

\bitem
\item [(1)] Using the expression (\ref{eq:BF^Ie0}) for the BF under the intrinsic prior and the explicit formulae of the marginal likelihoods
\eqref{MmN-0} and \eqref{eq:mult-m^IMe} one obtains
\ben
\label{eq:mult-BF^Ie0}
& & BF^I_{e0}(\bm{y}|t)=  \frac{\Gamma(t+rc)\Gamma(n+r)\Gamma(n+c)}{\Gamma(t+n+rc)\Gamma(t+r)\Gamma(t+c)} \nonumber \\
&\times&
\sum_{\{\bm{x}:\sum_{ij}x_{ij}=t\}} \binom{t}{\bm{x}} \frac{(\prod_{i}x_{i+}!)(\prod_{j}x_{+j}!)}{\prod_{i}y_{i+}!)(\prod_{j}y_{+j}!)}\times \frac{\prod_{i,j}(x_{ij}+y_{ij})!}{\prod_{i,j}x_{ij}!}.
\een
Note that the  sum in \eqref{eq:mult-BF^Ie0} is over all the  $r\times c$ tables with grand total $t$, which
 cannot be evaluated exactly in realistic settings: however it can be tackled  through a  Monte Carlo sum with an importance sampling algorithm.
 The candidate distribution we use is Multinomial with cell probabilities equal to the modified MLE estimates.

\item [(2)] The value of $BF^I_{ce}(\bm{y}|t)$ in equation (\ref{eq:BF^Ice}) can be computed  along the lines presented in Section \ref{sec:Hyp test prod binom}, namely through a direct sampling algorithm for evaluations under the intrinsic prior or  a Metropolis within Gibbs algorithm  under the intrinsic posterior.

\item [(3)]
Finally the value of    $BF^I_{c0}(\bm{y}|t)$  is computed as in (\ref{eq:BF^Ic0}).

\eitem

The details of the algorithms that we implemented are reported in Appendix.

\section{Simulations and real data analysis}
\label{sec:Applications}

In this section we evaluate features and performance of our approach through simulations  and apply our methodology to real datasets.
%
%

\subsection{Simulations  for the product binomial  model}
\label{subsec: Simulations}
We simulated 200 contingency tables for each of the scenarios described in Table \ref{tab:tabsim} characterized by a \emph{decreasing} pattern for the success probabilities $\pi_i$'s as the level $i$ of the row increases.
Thus the true model $M_c$ is constrained and
we can verify the ability of our method  to identify it.
The models under consideration   are
\begin{align*}
M_0:& \, \pi_1=\ldots =\pi_r \\
M_c:& \, \pi_1>\ldots >\pi_r \\
M_e:& \, (\pi_i, i=1,\dots,r )\in \{ [0,1]^r\setminus \{\pi_1 =\ldots =\pi_r\} \}.
\end{align*}

Our simulation settings under $M_c$ are based on  Cohen's   effect size (ES) measuring the separation between  two proportions or probabilities  \citep{Cohe:1992} expressed
  as absolute differences between the arcsine transformation of the probabilities.  We considered four types of ES:
  small ("S") if  ES=0.2; medium ("M") if ES=0.5;  large   ("L") if ES=0.8 and extra-large ("XL") if  ES$\geq 1$.

  We simulated product-binomial contingency tables of dimensions $2 \times 2$ and $3 \times 2$,  with true probabilities of success in decreasing order under the above four ES's, namely "S", "M", "L" and "XL", and within each we considered three scenarios.
  Since for $ 3 \times 2$ tables it is not possible to find triples of ordered probabilities with adjacent entries having ES="L" or ES="XL", the ES criterion refers to the smallest and the largest probability for each of the scenarios.

 For simplicity we let  the sample sizes be equal across rows, that is $n_i=n^*$, $i=1,\ldots,r$.
 For each given effect size,
  the sample size  $n^*$ is set in such a way that the test, with significance level 0.05, achieves
 a power of  0.80, conventionally regarded as adequate in most applications (for $3\times 2$ tables, significance incorporated  Bonferroni correction for multiple comparisons). The sample sizes can be found in Table 2 of \cite{Cohe:1992}; alternatively they can be   computed using the R package \textsf{pwr} \citep{R:pwr}.

%
{\renewcommand{\baselinestretch}{1}
\begin{table}[!ht]
\centering \caption{\protect\small \textit{
Simulation setting. Product binomial. Number of trials $n^*$ and true success probabilities $\pi_i$ for each scenario (\#) within effect size (ES).
}}
\label{tab:tabsim}
\begin{tabular}{|c|c|c|c|c|c|c|}
\hline
 & \multicolumn{3}{|c|}{table $2\times 2$} & \multicolumn{3}{|c|}{table $3\times 2$}\\
\hline
$ES$ & \# &$n^*$ & $(\pi_1;\pi_2)$ & \# & $n^*$ & $(\pi_1;\pi_2;\pi_3)$\\
\hline
S & 1 & 392 & $(0.10;0.05)$  & 1 & 441 & $(0.10;0.075;0.05)$  \\
  & 2 &     & $(0.50;0.40)$  & 2 &    & $(0.50; 0.45; 0.40) $  \\
  & 3 &     & $(0.95;0.90)$  & 3 &    & $(0.95; 0.92; 0.90) $  \\
\hline
M & 1 & 63 & $(0.30;0.10)$  & 1& 71 & $(0.30; 0.20; 0.10)$ \\
  & 2 &    & $(0.50;0.26)$ & 2 &    & $(0.50; 0.38  0.26)$  \\
  & 3 &    & $(0.90;0.70)$ & 3 &    & $(0.90; 0.80; 0.70)$  \\
\hline
L & 1 & 25 & $(0.60;0.22)$& 1 & 28 & $(0.60; 0.41; 0.22)$  \\
  & 2 &    & $(0.80;0.42)$ &2 &    & $(0.80; 0.61; 0.42)$  \\
  & 3 &    & $(0.90;0.56)$ &3 &    &$(0.90; 0.73; 0.56)$  \\
     \hline
XL & 1 & 13 & $(0.60;0.15)$ & 1 & 15& $(0.60;0.30;0.15)$  \\
   & 2 &   & $(0.80;0.20)$ & 2 &   & $(0.80;0.50;0.20)$  \\
   & 3 &   & $(0.90;0.25)$ & 3 &   & $(0.90;0.60;0.25)$  \\
\hline
\end{tabular}
\end{table}
}
Table \ref{tab:tabsim} reports, for each scenario (\#) within effect-size (ES),   the number of trials
$n^*$ and true success probabilities $\pi_i$.
With regard to the specification  of the intrinsic prior,  we let the training sample size $t^*$ vary from $0$ to $n^*$; for simplicity of exposition however,  only results for a few selected values are reported, namely those corresponding to $q=0,0.25,0.5,0.75,1$, where
$q=t^*/n^*$ is the ratio between the training and the actual sample size.



Before proceeding with the discussion of the results, we provide further insights on  the nature of the intrinsic prior and the likelihood function across different values of ES for  $2\times 2$ contingency tables.
Figure \ref{contour-lines} plots the contour lines of the intrinsic prior densities for selected values of $q$:  0.25 (black),  0.5 (green) 0.75 (yellow),
together with those of the (normalized) likelihood function based on $n=100$ simulated observations (red).

Two features emerge from Figure \ref{contour-lines}.
The dependence of the intrinsic prior on the training sample size $t^*$ (equivalently $q$), a feature  already described in Fig.\ref{intr-prior}, is apparent also in this case. As $q$ increases the prior mass progressively concentrates around the space characterizing $M_0$; notice however that the intrinsic prior piles up much more mass in the neighborhood of the corners $ \{(0,0),  (1,1) \}$ than along any of the  points on the line $\pi_1=\pi_2$. 
Accordingly only the two intrinsic priors corresponding to $q=0.25$ and $q=0.5$ have some traceable contours outside the two corners.

With regard to the likelihood, if the effect size is small, the data are in broad agreement with model $M_0$,  and the contour lines of the (normalized) likelihood do overlap with the prior contours.  As the effect size becomes larger (raising  to "M", "L" or "XL") the bulk of the likelihood moves away  from that of the  prior, because the data are progressively departing from the null model. Notice however that, in the area wherein it is concentrated,  the likelihood, has much higher values than the prior; accordingly the marginal (integrated) likelihood for model $M_c$ will be appreciable allowing the constrained model to compete strongly against $M_0$ and $M_e$.

{\renewcommand{\baselinestretch}{0.9}
\begin{figure}
\subfloat[Small effect size]{\includegraphics[width = 3in]{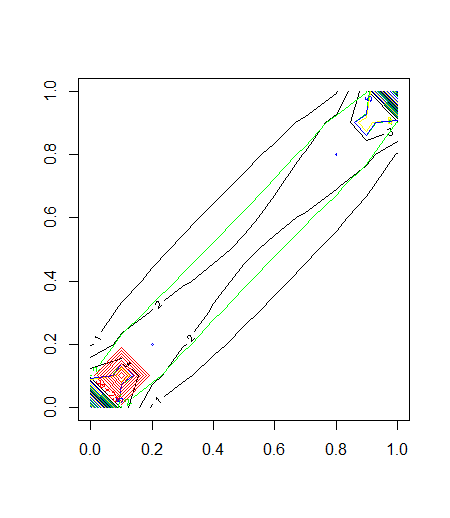}}
\subfloat[Medium effect size]{\includegraphics[width = 3in]{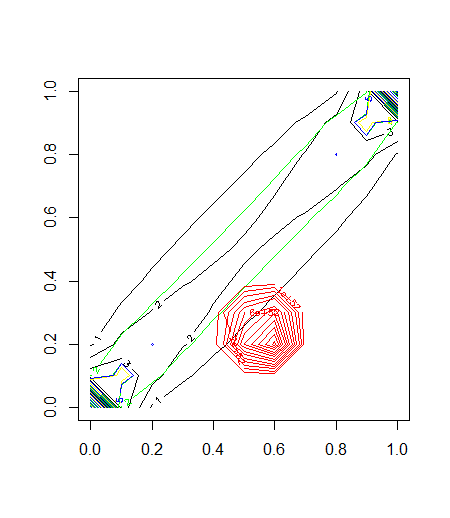}}\\
\subfloat[Large effect size]{\includegraphics[width = 3in]{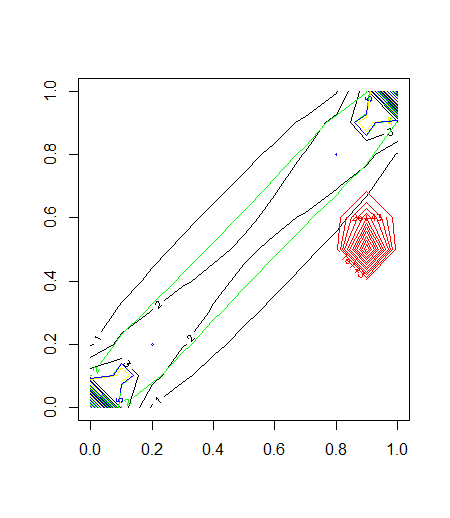}}
\subfloat[XLarge effect size]{\includegraphics[width = 3in]{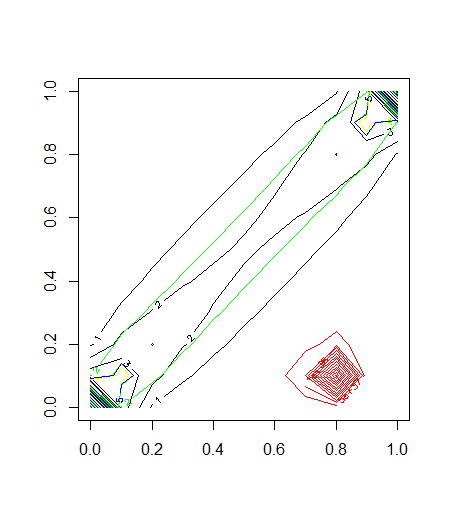}}
\caption{\protect\small \textit{
$2\times 2$ table under product binomial ($n_1=n_2=n=100$) with values of effect size "S", "M", "L", "XL".
Contour plots of intrinsic priors: $q=0.25$ (black),  $q=0.5$ (green), $q= 0.75$ (yellow), and
(normalized) likelihood (red).
}}
\label{contour-lines}
\end{figure}
}


Tables \ref{tab:tab2x2} and \ref{tab:tab3x2} report the median (across the 200 simulations) of the posterior probabilities of the null model, $\prob^I(M_0 |t,\bm{y})$,  and of  the correct model, $\prob^I(M_c |t, \bm{y})$, for selected values of the  fraction $q$ of the training sample size.
 Both for contingency tables $2\times 2$ and $3\times 2$  two sets of model comparison were considered:  $\{M_0, M_c, M_e \}$ and  $\{M_0,M_c \}$. Within each set the prior on models is taken to be uniform.


Consider first Table \ref{tab:tab2x2}.
If the model set is  $\{M_0, M_c, M_e \}$ , the posterior probability of the true model $M_c$
is in the range 50\%-99\% across all scenarios, with values increasing as the effect size becomes larger.
In particular,  posterior probabilities between 50\% and 60\% occur only when the ES is either Small or Medium, whereas they are never below the threshold 77\% when the ES is Large or XLarge.
 If however the comparison is restricted to the pair  $\{M_0, M_c\}$
the above range drastically shrinks to 83\%-99\% with a similar behavior with respect to increasing levels of  ES. There is also a remarkable robustness  to varying levels of $q$. 
Broadly similar considerations apply to Table \ref{tab:tab3x2};  we omit details in the interest of brevity.
In conclusion our method
is able to identify the true model  even when the effect size is small (e.g. using a conventional threshold of 50\%),  exhibits very limited sensitivity to the  size of the imaginary sample used to construct the intrinsic prior, and behaves sensibly with regard to increasing levels of effect size.

%
%
%
%
%
%
%
%
%

\newpage

{\small
	{\renewcommand{\baselinestretch}{0.9}
		\begin{table}[h]
			\centering \caption{\protect\small \textit{Simulation study for $2\times 2$ contingency tables. Median of the posterior model probabilities as a function of $q$ for distinct model comparison sets. Notice that in the labels of the columns the notation of the posterior model probability is  $\prob^I_{t,\bm{y}}(M)$  instead of  $\prob^I(M|t,\bm{y})$.
			}}
			\label{tab:tab2x2}
			\begin{tabular}{|c|c|c|c||c|c|c||c|c|c|}
				\hline
				& \multicolumn{3}{|c|}{\# S1} & \multicolumn{3}{|c|}{\# S2} & \multicolumn{3}{|c|}{\# S3} \\
				\hline
				&  \multicolumn{2}{|c|}{$M_0$-$M_c$-$M_e$} & \multicolumn{1}{|c||}{$M_0$-$M_c$} &  \multicolumn{2}{|c|}{$M_0$-$M_c$-$M_e$} & \multicolumn{1}{|c||}{$M_0$-$M_c$} &  \multicolumn{2}{|c|}{$M_0$-$M_c$-$M_e$} & \multicolumn{1}{|c||}{$M_0$-$M_c$} \\
				\hline
				$q$ &
				{\tiny $\prob^I_{t,\bm{y}}(M_0)$ }& {\tiny $\prob^I_{t,\bm{y}}(M_c)$} &  {\tiny $\prob^I_{t,\bm{y}}(M_c)$ }
				& {\tiny $\prob^I_{t,\bm{y}}(M_0)$}& {\tiny $\prob^I_{t,\bm{y}}(M_c)$} & {\tiny $\prob^I_{t,\bm{y}}(M_c)$ } &
				{\tiny $\prob^I_{t,\bm{y}}(M_0)$ }& {\tiny $\prob^I_{t,\bm{y}}(M_c)$} & {\tiny $\prob^I_{t,\bm{y}}(M_c)$ } \\
				\hline
				0   & 0.035 & 0.593 &0.862 &0.040 &0.565 &0.892 &0.037 &0.587& 0.892\\
				0.25 & 0.036 & 0.682 &0.847 &0.035 &0.593 &0.894 &0.036 &0.597 &0.892\\
				0.5  & 0.038 & 0.581 &0.832 &0.035 &0.609 &0.894 &0.039 &0.521 &0.891\\
				0.75 & 0.055 & 0.571 &0.900 &0.052 &0.601 &0.892 &0.038 &0.548 &0.891\\
				1 & 0.067 & 0.504 &0.873 &0.041 &0.616 &0.894 &0.039 &0.567&0.890\\
				\hline
				\hline
				& \multicolumn{3}{|c|}{\# M1} & \multicolumn{3}{|c|}{\# M2} & \multicolumn{3}{|c|}{\# M3} \\
				\hline
				&  \multicolumn{2}{|c|}{$M_0$-$M_c$-$M_e$} & \multicolumn{1}{|c||}{$M_0$-$M_c$} &  \multicolumn{2}{|c|}{$M_0$-$M_c$-$M_e$} & \multicolumn{1}{|c||}{$M_0$-$M_c$} &  \multicolumn{2}{|c|}{$M_0$-$M_c$-$M_e$} & \multicolumn{1}{|c||}{$M_0$-$M_c$} \\
				\hline
				$q$ &
				{\tiny $\prob^I_{t,\bm{y}}(M_0)$ }& {\tiny $\prob^I_{t,\bm{y}}(M_c)$} &  {\tiny $\prob^I_{t,\bm{y}}(M_c)$ }
				& {\tiny $\prob^I_{t,\bm{y}}(M_0)$}& {\tiny $\prob^I_{t,\bm{y}}(M_c)$} & {\tiny $\prob^I_{t,\bm{y}}(M_c)$ } &
				{\tiny $\prob^I_{t,\bm{y}}(M_0)$ }& {\tiny $\prob^I_{t,\bm{y}}(M_c)$} & {\tiny $\prob^I_{t,\bm{y}}(M_c)$ } \\
				\hline
				0   &0.026 & 0.571 & 0.842 & 0.028 & 0.591 & 0.863 & 0.030 &0.600& 0.887\\
				0.25 &0.027 & 0.608 & 0.857 & 0.026 &0.587 & 0.864 & 0.028& 0.582 &	0.896\\
				0.5  & 0.028 & 0.626 & 0.859 & 0.029 &0.607 & 0.865 & 0.025 & 0.606 &0.896\\
				0.75 & 0.022 & 0.594 & 0.860& 0.026 &0.617 & 0.866 & 0.026 & 0.600 & 0.894\\
				1    & 0.022 & 0.541 & 0.821& 0.028 &0.610 & 0.865 & 0.029 & 0.605 & 0.895\\
				\hline
				\hline
				& \multicolumn{3}{|c|}{\# L1} & \multicolumn{3}{|c|}{\# L2} & \multicolumn{3}{|c|}{\# L3} \\
				\hline
				&  \multicolumn{2}{|c|}{$M_0$-$M_c$-$M_e$} & \multicolumn{1}{|c||}{$M_0$-$M_c$} &  \multicolumn{2}{|c|}{$M_0$-$M_c$-$M_e$} & \multicolumn{1}{|c||}{$M_0$-$M_c$} &  \multicolumn{2}{|c|}{$M_0$-$M_c$-$M_e$} & \multicolumn{1}{|c||}{$M_0$-$M_c$} \\
				\hline
				$q$ &
				{\tiny $\prob^I_{t,\bm{y}}(M_0)$ }& {\tiny $\prob^I_{t,\bm{y}}(M_c)$} &  {\tiny $\prob^I_{t,\bm{y}}(M_c)$ }
				& {\tiny $\prob^I_{t,\bm{y}}(M_0)$}& {\tiny $\prob^I_{t,\bm{y}}(M_c)$} & {\tiny $\prob^I_{t,\bm{y}}(M_c)$ } &
				{\tiny $\prob^I_{t,\bm{y}}(M_0)$ }& {\tiny $\prob^I_{t,\bm{y}}(M_c)$} & {\tiny $\prob^I_{t,\bm{y}}(M_c)$ } \\
				\hline
				0   &  0.049 & 0.690 & 0.934 & 0.045 & 0.682 & 0.922 & 0.048 &0.682	& 0.954\\
				0.25 & 0.040 & 0.682 & 0.935 & 0.042 &  0.728 & 0.928 & 0.042 & 0.656 &	0.952\\
				0.5  & 0.041 & 0.699 & 0.936 & 0.042 &  0.708 & 0.955 & 0.043 & 0.720 & 0.951\\
				0.75 & 0.043 & 0.691 & 0.950& 0.044 &  0.715 & 0.955 & 0.044 & 0.716 & 0.949\\
				1    & 0.040 & 0.699 & 0.938& 0.041 &  0.715 & 0.940 & 0.045 & 0.716 & 0.949\\
				\hline
				\hline
				& \multicolumn{3}{|c|}{\# XL1} & \multicolumn{3}{|c|}{\# XL2} & \multicolumn{3}{|c|}{\# XL3} \\
				\hline
				&  \multicolumn{2}{|c|}{$M_0$-$M_c$-$M_e$} & \multicolumn{1}{|c||}{$M_0$-$M_c$} &  \multicolumn{2}{|c|}{$M_0$-$M_c$-$M_e$} & \multicolumn{1}{|c||}{$M_0$-$M_c$} &  \multicolumn{2}{|c|}{$M_0$-$M_c$-$M_e$} & \multicolumn{1}{|c||}{$M_0$-$M_c$} \\
				\hline
				$q$ &
				{\tiny $\prob^I_{t,\bm{y}}(M_0)$ }& {\tiny $\prob^I_{t,\bm{y}}(M_c)$} &  {\tiny $\prob^I_{t,\bm{y}}(M_c)$ }
				& {\tiny $\prob^I_{t,\bm{y}}(M_0)$}& {\tiny $\prob^I_{t,\bm{y}}(M_c)$} & {\tiny $\prob^I_{t,\bm{y}}(M_c)$ } &
				{\tiny $\prob^I_{t,\bm{y}}(M_0)$ }& {\tiny $\prob^I_{t,\bm{y}}(M_c)$} & {\tiny $\prob^I_{t,\bm{y}}(M_c)$ } \\
				\hline
				0    & 0.059 & 0.862 & 0.995 & 0.074 & 0.867 & 0.995 & 0.069 & 0.895 & 0.997\\
				0.25 & 0.061 & 0.856 & 0.992 & 0.070 & 0.862 & 0.986 & 0.065 & 0.898 & 0.991\\
				0.5  & 0.056 & 0.858 & 0.991 & 0.071 & 0.862 & 0.981 & 0.070 & 0.900 & 0.990\\
				0.75 & 0.072 & 0.858 & 0.989 & 0.069 & 0.863 & 0.975 & 0.070 & 0.900 & 0.993\\
				1    & 0.075 & 0.858 & 0.989 & 0.071 & 0.863 & 0.970 & 0.073 & 0.899 & 0.999\\
				\hline	
			\end{tabular}
		\end{table}
	}
}
%

\newpage

{\small
		{\renewcommand{\baselinestretch}{0.9}
	\begin{table}[thb]
		\centering \caption{\protect\small \textit{Simulation study for $3\times 2$ contingency tables. Median of the posterior model probabilities as a function of $q$ for distinct model comparison sets. Notice that in the labels of the columns the notation of the posterior model probability is  $\prob^I_{t,\bm{y}}(M)$  instead of  $\prob^I(M|t,\bm{y})$.
				}}
		\label{tab:tab3x2}
		\begin{tabular}{|c|c|c|c||c|c|c||c|c|c|}
			\hline
			& \multicolumn{3}{|c|}{\# S1} & \multicolumn{3}{|c|}{\# S2} & \multicolumn{3}{|c|}{\# S3} \\
			\hline
		  &  \multicolumn{2}{|c|}{$M_0$-$M_c$-$M_e$} & \multicolumn{1}{|c||}{$M_0$-$M_c$} &  \multicolumn{2}{|c|}{$M_0$-$M_c$-$M_e$} & \multicolumn{1}{|c||}{$M_0$-$M_c$} &  \multicolumn{2}{|c|}{$M_0$-$M_c$-$M_e$} & \multicolumn{1}{|c||}{$M_0$-$M_c$} \\
			\hline
			$q$ &
			{\tiny $\prob^I_{t,\bm{y}}(M_0)$ }& {\tiny $\prob^I_{t,\bm{y}}(M_c)$} &  {\tiny $\prob^I_{t,\bm{y}}(M_c)$ }
			& {\tiny $\prob^I_{t,\bm{y}}(M_0)$}& {\tiny $\prob^I_{t,\bm{y}}(M_c)$} & {\tiny $\prob^I_{t,\bm{y}}(M_c)$ } &
			{\tiny $\prob^I_{t,\bm{y}}(M_0)$ }& {\tiny $\prob^I_{t,\bm{y}}(M_c)$} & {\tiny $\prob^I_{t,\bm{y}}(M_c)$ } \\
			\hline
			0    & 0.025 & 0.565 &0.611 &0.021 &0.569 &0.655 &0.022 &0.557 &0.656\\
			0.25 & 0.023 & 0.565 &0.596 &0.021 &0.569 &0.616 &0.024 &0.565 &0.646\\
			0.5  & 0.018 & 0.570 &0.613 &0.022 &0.577 &0.682 &0.024 &0.568 &0.666\\
			0.75 & 0.021 & 0.572 &0.630 &0.022 &0.582 &0.660 &0.024 &0.580 &0.669\\
			1    & 0.031 & 0.574 &0.654 &0.030 &0.587 &0.679 &0.025 &0.575 &0.674\\
			\hline
			\hline
		& \multicolumn{3}{|c|}{\# M1} & \multicolumn{3}{|c|}{\# M2} & \multicolumn{3}{|c|}{\# M3} \\
		\hline
		&  \multicolumn{2}{|c|}{$M_0$-$M_c$-$M_e$} & \multicolumn{1}{|c||}{$M_0$-$M_c$} &  \multicolumn{2}{|c|}{$M_0$-$M_c$-$M_e$} & \multicolumn{1}{|c||}{$M_0$-$M_c$} &  \multicolumn{2}{|c|}{$M_0$-$M_c$-$M_e$} & \multicolumn{1}{|c||}{$M_0$-$M_c$} \\
		\hline
		$q$ &
		{\tiny $\prob^I_{t,\bm{y}}(M_0)$ }& {\tiny $\prob^I_{t,\bm{y}}(M_c)$} &  {\tiny $\prob^I_{t,\bm{y}}(M_c)$ }
		& {\tiny $\prob^I_{t,\bm{y}}(M_0)$}& {\tiny $\prob^I_{t,\bm{y}}(M_c)$} & {\tiny $\prob^I_{t,\bm{y}}(M_c)$ } &
		{\tiny $\prob^I_{t,\bm{y}}(M_0)$ }& {\tiny $\prob^I_{t,\bm{y}}(M_c)$} & {\tiny $\prob^I_{t,\bm{y}}(M_c)$ } \\
		\hline
		0   &0.021 & 0.576 & 0.693 & 0.029 & 0.588 & 0.769 & 0.025 &0.558& 0.703\\
		0.25 &0.021 & 0.573 & 0.712 & 0.029 &0.587 & 0.817 & 0.025& 0.558 &	0.759\\
		0.5  & 0.023 & 0.576 & 0.770 & 0.029 &0.588 & 0.798 & 0.025 & 0.556&0.779\\
		0.75 & 0.022 & 0.579 & 0.743& 0.029 &0.588 & 0.794 & 0.026 & 0.559 & 0.789\\
		1    & 0.022 & 0.580 & 0.763& 0.028 &0.588 & 0.796 & 0.025 & 0.560 & 0.783\\
		\hline
		\hline
	& \multicolumn{3}{|c|}{\# L1} & \multicolumn{3}{|c|}{\# L2} & \multicolumn{3}{|c|}{\# L3} \\
	\hline
	&  \multicolumn{2}{|c|}{$M_0$-$M_c$-$M_e$} & \multicolumn{1}{|c||}{$M_0$-$M_c$} &  \multicolumn{2}{|c|}{$M_0$-$M_c$-$M_e$} & \multicolumn{1}{|c||}{$M_0$-$M_c$} &  \multicolumn{2}{|c|}{$M_0$-$M_c$-$M_e$} & \multicolumn{1}{|c||}{$M_0$-$M_c$} \\
	\hline
	$q$ &
	{\tiny $\prob^I_{t,\bm{y}}(M_0)$ }& {\tiny $\prob^I_{t,\bm{y}}(M_c)$} &  {\tiny $\prob^I_{t,\bm{y}}(M_c)$ }
	& {\tiny $\prob^I_{t,\bm{y}}(M_0)$}& {\tiny $\prob^I_{t,\bm{y}}(M_c)$} & {\tiny $\prob^I_{t,\bm{y}}(M_c)$ } &
	{\tiny $\prob^I_{t,\bm{y}}(M_0)$ }& {\tiny $\prob^I_{t,\bm{y}}(M_c)$} & {\tiny $\prob^I_{t,\bm{y}}(M_c)$ } \\
	\hline
	0   &  0.033 & 0.770 & 0.980 & 0.032 & 0.777 & 0.976 & 0.029 &0.818	& 0.973\\
	0.25 & 0.031 & 0.768 & 0.973 & 0.035 &  0.788 & 0.978 & 0.030 & 0.818 &	0.972\\
	0.5  & 0.033 & 0.810 & 0.979 & 0.035 &  0.781 & 0.974 & 0.030 & 0.812 & 0.972\\
	0.75 & 0.033 & 0.808 & 0.979& 0.035 &  0.789 & 0.974 & 0.031 & 0.816 & 0.972\\
	1    & 0.033 & 0.801 & 0.979& 0.036 &  0.788 & 0.973 & 0.030 & 0.811 & 0.972\\
	\hline
	\hline
		& \multicolumn{3}{|c|}{\# XL1} & \multicolumn{3}{|c|}{\# XL2} & \multicolumn{3}{|c|}{\# XL3} \\
	\hline
	&  \multicolumn{2}{|c|}{$M_0$-$M_c$-$M_e$} & \multicolumn{1}{|c||}{$M_0$-$M_c$} &  \multicolumn{2}{|c|}{$M_0$-$M_c$-$M_e$} & \multicolumn{1}{|c||}{$M_0$-$M_c$} &  \multicolumn{2}{|c|}{$M_0$-$M_c$-$M_e$} & \multicolumn{1}{|c||}{$M_0$-$M_c$} \\
	\hline
	$q$ &
	{\tiny $\prob^I_{t,\bm{y}}(M_0)$ }& {\tiny $\prob^I_{t,\bm{y}}(M_c)$} &  {\tiny $\prob^I_{t,\bm{y}}(M_c)$ }
	& {\tiny $\prob^I_{t,\bm{y}}(M_0)$}& {\tiny $\prob^I_{t,\bm{y}}(M_c)$} & {\tiny $\prob^I_{t,\bm{y}}(M_c)$ } &
	{\tiny $\prob^I_{t,\bm{y}}(M_0)$ }& {\tiny $\prob^I_{t,\bm{y}}(M_c)$} & {\tiny $\prob^I_{t,\bm{y}}(M_c)$ } \\
	\hline
	0    & 0.075 & 0.862 & 0.993 & 0.045 & 0.873 & 0.997 & 0.065 & 0.883 & 0.998\\
	0.25 & 0.081 & 0.857 & 0.998 & 0.045 & 0.877 & 0.999 & 0.069 & 0.890 & 0.999\\
	0.5  & 0.052 & 0.864 & 0.997 & 0.045 & 0.876 & 0.999 & 0.068 & 0.889 & 0.998\\
	0.75 & 0.057 & 0.867 & 0.997 & 0.046 & 0.876 & 0.999 & 0.068 & 0.889 & 0.999\\
	1    & 0.065 & 0.868 & 0.997 & 0.045 & 0.878 & 0.999 & 0.067 & 0.889 & 0.999\\
	\hline	
	\end{tabular}
	\end{table}
}
}
%

\subsection{Real data analyses}
\label{subsec: Real data}
In this subsection we apply  our method to  real datasets and compare our results with previously analyzed studies.  One aspect which we further consider is the robustness of our conclusions to the choice of the hyper-parameter $q$ which represents the fraction of the training sample size,  relative to the actual sample size, which is used to construct the intrinsic prior.
Assuming lack of prior information,  all models under consideration are given \emph{a priori} the same probability. Different prior model probabilities can be easily accommodated within our framework.

\subsubsection{Product binomial model}

\paragraph{Trauma due to subarachnoid hemorrhage.
}

We return  to Table 1 of Section 3 which reports    the response to different  treatments  of patients who experienced trauma due to subarachnoid hemorrhage  \citep{Agre:Coull:2002}. This is a $4 \times 2$ contingency table whose columns are the response categories (\lq \lq dead\rq \rq{} or \lq \lq not dead \rq \rq{}) while  the rows contain three ordered levels of medication dose plus a control group. The objective of the study is to determine whether a more favorable outcome tends to occur as the dose increases. Using the notation of Section \ref{sec:IEApp}   there are three possible models that we can consider, namely
\begin{align*}
M_0:\, & \pi_1=\pi_2=\pi_3=\pi_4 \\
M_c: \, & \pi_1>\pi_2>\pi_3>\pi_4 \\
M_e: \, & (\pi_i, i=1,\dots,4 )\in \{ [0,1]^4\setminus \{\pi_1 =\ldots =\pi_4\} \}.
\end{align*}
\citet{Agre:Coull:2002} analyzed these data using a frequentist approach. Specifically, they  tested the null model of equal probabilities $M_0$ against that of  ordered alternatives $M_c$ using the large-sample chi-bar squared distribution, and obtained  a $p$-value equal to 0.095, so that the null model cannot be rejected using default settings. However they correctly point out that this result does not  enable one to conclude how strong is the  evidence in favor of the null. The latter  instead is available using our approach.
%

%
{\renewcommand{\baselinestretch}{0.9}
\begin{table}[h]
\centering \caption{\protect\small \textit{Trauma due to subarachnoid hemorrhage. Bayes  factors as a function of $q$.
}} \label{tab:resAgr-Coull-BF}
\begin{tabular}{|c|c|c|c|}
\hline
 $q$ & $BF^I_{e0}$ & $BF^I_{ce}$ & $BF^I_{c0}$  \\
\hline
 0    &  12.965 & 1.017 & 13.189 \\
 0.25 & 144.197 & 2.172 & 313.259 \\
 0.5 &  147.798 & 2.752 & 406.793 \\
 0.75&  123.942  & 3.336 & 413.453 \\
 1   &  93.672   & 3.737  & 350.017 \\
\hline
\end{tabular}
\end{table}
}
{\renewcommand{\baselinestretch}{0.9}
	\begin{table}[h]
		\centering \caption{\protect\small \textit{Trauma due to subarachnoid hemorrhage. Posterior model probabilities as a function of $q$ for distinct  model comparison sets. }} \label{tab:resAgr-Coull-PP}
		\begin{tabular}{|c|c|c|c|c|c|c|c|c|}
			\hline
			&  \multicolumn{2}{|c|}{$M_0$ - $M_e$} & \multicolumn{2}{|c|}{$M_0$ - $M_c$} & \multicolumn{3}{|c|}{$M_0$ - $M_c$ - $M_e$} \\
			\hline
			$q$ & {\scriptsize $\prob^I(M_0 |t, \bm{y})$ }& {\scriptsize$\prob^I(M_e |t, \bm{y})$} & {\scriptsize $\prob^I(M_0 |t, \bm{y})$} & {\scriptsize $\prob^I(M_c |t, \bm{y})$ }& {\scriptsize $\prob^I(M_0 |t, \bm{y})$} & {\scriptsize $\prob^I(M_c |t, \bm{y})$} & {\scriptsize $\prob^I(M_e |t, \bm{y})$}\\
			\hline
			0 & 0.072 & 0.923 & 0.070& 0.930 & 0.037& 0.504 & 0.459\\
			0.25 & 0.007 & 0.993 & 0.003 & 0.997& 0.002 & 0.685& 0.313\\
			0.5 & 0.007 & 0.993 & 0.002 & 0.998& 0.002 & 0.733 & 0.265\\
			0.75 & 0.008 & 0.992 & 0.002 & 0.998 & 0.002 & 0.769& 0.229\\
			1 & 0.011 & 0.990 & 0.003 & 0.997 & 0.002 & 0.789 & 0.209\\
			\hline
		\end{tabular}
	\end{table}
}

\black
 Table \ref{tab:resAgr-Coull-BF} reports,  for selected values of $q$,  the  Bayes factors $BF^I_{e0}$,  $BF^I_{ce}$ and  $BF^I_{c0}$ (the last one being of course a function of the former two). It appears that both the unconstrained and the constrained model are strongly  supported by the data relative to the null model of independence with values of $BF^I_{e0}$ over 100 and those of  $BF^I_{c0}$ over 300 for $q \geq 0.25$.
  In other words  the strength of evidence \citep{Scho:Wage:2018} against the null is \emph{extreme} whether the comparison is made against the unconstrained or the constrained model.
 Additionally the Bayes factor for comparing $M_c$ against $M_e$ suggests values greater than 1 and extending beyond 3.5. Although this represents only \emph{anecdotal}, or at most \emph{moderate} evidence,  in favor of $M_c$,  it nevertheless indicates that $M_c$ is somewhat better supported by the data than $M_e$.
 Table \ref{tab:resAgr-Coull-PP} allows a finer appreciation of the main features of our analysis, as it
 reports the posterior probability of each model separately for each
  of the three  sets of model comparison, namely
   $\{ M_0,M_e \}$,  $\{ M_0,M_c \}$ and  $\{ M_0,M_c,M_e \}$.
It appears that the evidence in favor of the null,  $Pr^I(M_0 |t, \bf{y})$,  is very small, its value never exceeding $7\%$ while being most of the time a tenth of the above or lower. Accordingly the  constrained model $M_c$  receives a very high posterior probability (above 90\%) when the comparison is restricted to $\{ M_0,M_c \}$; this value somewhat diminishes  (being around  70\%) when also the  unconstrained model is taken into consideration.
   We therefore conclude that evidence in favor of the constrained model $M_c$ is strong  and that this result is \emph{robust} to variations in $q$.

   We highlight
    the fact that the constrained model $M_c$ and the unconstrained model $M_e$ have the \emph{ same dimension}.
     Nevertheless $M_c$ is \emph{nested} into, and so \emph{less complex} than,   $M_e$
     because of its smaller parameter space.
    Interestingly,  $M_c$ receives a much higher  posterior probability than $M_e$ as it is apparent from scenario $\{M_0, M_c, M_e  \}$,   at least for  $q \geq 0.25$. This occurs because of the more complex models are penalized due to  Ockham's razor \citep{Jeff:Berg:1992}.
      We therefore conclude that  not only is the null model of independence to be discarded, but there is clear evidence in favor of the constrained model.

\subsubsection{Multinomial model}

\paragraph{Surgical methods for ulcer treatment.
}

\citet{Efron:1996} analyzed data coming   from a multicenter trial whose objective was to establish whether a new surgical method (Treatment) for ulcer was superior to an older one (Control) with regard to reducing   recurrent bleeding.
 The data refer to 41 hospitals. For each hospital  a $2\times 2$ contingency table  summarizes the results.
Each table  is presented as $(a,b;c,d)$,  where $(a, b)$ are  the number of occurrences and non-occurrences
for the Treatment,   while $(c, d)$  are the corresponding values  for the Control; here
 occurrence refers to recurrent bleeding.

\citet{Case:More:2009} tested in each contingency table  independence between  occurrence and method of surgery (model $M_0$) against an unconstrained alternative ($M_e$):
\begin{eqnarray}
\label{eq:multinomial-M0-M1}
M_0&:& \pi_{ij}=\theta_i \cdot \eta_j;\, \forall i,j\in\{1,2\} \, (0<\theta_1<1; 0<\eta_1<1; \,   \theta_1+\theta_2=1; \,\eta_1+\eta_2=1) \nonumber \\
M_e&:&   0<\pi_{ij}<1;\, (\pi_{11}+\pi_{12}+\pi_{21}+\pi_{22}=1).
\end{eqnarray}
Letting $\theta_1=\theta$ and $\eta_1=\eta$,  they used the following default priors
\begin{eqnarray}
\label{casella-moreno-priors}
p^N(\theta,\eta |M_0)&=&p^N(\theta|M_0)p^N(\eta|M_0)=Unif(\theta; 0,1)Unif(\eta; 0,1) \nonumber \\
p^N(\bm{\pi} |M_e)&=&Di(\bm{\pi}|1,1,1,1).
\end{eqnarray}
Next they constructed an intrinsic prior under $M_e$ letting the training sample size $t$ range over the set $0, \ldots,n$.
Their results are reported
in Table \ref{tab:Case:More} for five selected hospitals  arranged according to increasing $p$-values.
\black
Although the posterior probability of the null model is conceptually quite different from the $p$-value
\citep{Wass:Laza:2016},
one can see that it
generally increases with the $p$-value correctly reporting higher evidence in favor of the null.
However only for one hospital (\# 18) does the posterior probability of the null exceeds the 0.5 threshold (when $q=0)$, and even in this case  the result is not robust because it goes below this value when $q=1$.
Based on the intrinsic  analysis they conclude that for none of these  hospitals there exists a robust support for the  null hypothesis of independence of surgery and occurrence.


{\renewcommand{\baselinestretch}{0.9}
\begin{table}[h]
\centering \caption{\protect\small \textit{Selected contingency tables from \citet{Efron:1996}.    Posterior probability of the null model based on the intrinsic prior approach  for  $t=0$  and $t=n$. (The notation (a,b;c,d) denotes a $2\times 2$ table with first row (a,b) reporting adverse occurences and non occurrences for the new surgery and similarly  for the second row (c,d) which refers to the old surgery. $n=a+b+c+d$). Table adapted from \citet[Table 2]{Case:More:2009}}} \label{tab:Case:More}
\begin{tabular}{|c|c|c|c|c|}
\hline
Hospital number &  Data & $p$-value  &\multicolumn{2}{|c|}{ $\prob^I(M_0 |t, \bm{y})$} \\
\hline
     &           &        &  $t=0$   & $t=n$ \\
\cline{4-5}
 34 & (20,0;18,5) & 0.051& 0.215  & 0.215\\
 \hline
 1  & (8,7;2,11)& 0.054 & 0.170 & 0.253\\
 \hline
 38 &  (43,4;14,5)& 0.106 & 0.395 & 0.340\\
 \hline
 18 & (30,1;23,4) & 0.173 & 0.551 & 0.406\\
  \hline
 16 & (7,4;4,6) & 0.395& 0.451 & 0.497\\
\hline
\end{tabular}
\end{table}
}

A natural hypothesis underlying  Efron's data is that the new surgery is superior to the old one, i.e. the probability of occurrence within Treatment is lower than the corresponding probability under Control. However this feature is not taken into account in the previous analysis.
Accordingly, we reanalyze Efron's tables  explicitly accounting for this  hypothesis which we can write as:
\be
M_c: \frac{\pi_{11}}{\pi_{11}+\pi_{12}}<\frac{\pi_{21}}{\pi_{21}+\pi_{22}}, \ee
equivalently $Pr(\mbox{occurrence}|\mbox{Treament})<Pr(\mbox{occurrence}|\mbox{Control})$.

Tables \ref{tab:resEfron-BF} and \ref{tab:resEfron-PP} present our results which are obtained using default priors specified in
\eqref{casella-moreno-priors}.

{\renewcommand{\baselinestretch}{0.9}
	\begin{table}[h]
		\centering \caption{\protect\small \textit{
Selected contingency tables from \citet{Efron:1996}.
Bayes factors appearing in formula (\ref{eq:BF^Ic0}) as a function of $q$.
}}
 \label{tab:resEfron-BF}
		\begin{tabular}{|c||c|c|c||c|c|c||c|c|c|}
			\hline
			& \multicolumn{3}{|c||}{Table 34} & \multicolumn{3}{|c||}{Table 1} & \multicolumn{3}{|c|}{Table 38} \\
			\hline
			$q$ & $BF^I_{e0}$ & $BF^I_{ce}$ & $BF^I_{c0}$ & $BF^I_{e0}$ & $BF^I_{ce}$ & $BF^I_{c0}$& $BF^I_{e0}$ & $BF^I_{ce}$ & $BF^I_{c0}$ \\
			\hline
			0    &3.648 &	0.993& 3.624 & 4.892& 0.996 & 4.875& 1.529& 	0.987&	1.509 \\
			0.25 &4.812& 0.593& 2.852 & 4.003& 0.559 & 2.241 & 2.052 & 0.478& 0.982  \\
			0.5  &4.758& 0.395 &1.879 & 3.438& 0.379 &	1.306& 2.064 & 0.337 & 0.695  \\
			0.75 &4.392 & 0.277& 1.218 & 3.148& 0.278& 0.875 & 2.002& 0.255& 0.511  \\
			1    &4.054& 0.201& 0.816 &  2.882& 0.208 & 0.601 & 1.913 &	0.192& 0.368  \\
			\hline
			& \multicolumn{3}{|c||}{Table 18} & \multicolumn{3}{|c|}{Table 16} & \multicolumn{3}{}{} \\
			\cline{1-7}
			$q$ & $BF^I_{e0}$ & $BF^I_{ce}$ & $BF^I_{c0}$ & $BF^I_{e0}$ & $BF^I_{ce}$ & $BF^I_{c0}$  & \multicolumn{3}{}{} \\
			\cline{1-7}
			0    &0.815& 1.007& 0.821 &1.217 &0.998& 1.216 \\
			0.25 &1.305& 0.683& 0.891 &1.071 & 0.765& 0.819 \\
			0.5  &1.439& 0.506& 0.727 &0.996 & 0.676 & 0.674 \\
			0.75 &1.46 & 0.425 & 0.620 & 0.986 & 0.588 & 0.579 \\
			1    &1.439& 0.351 & 0.505 &1.000 & 0.525 & 0.525\\
			\cline{1-7}
			
		\end{tabular}
	\end{table}
}
%
{\renewcommand{\baselinestretch}{0.9}
	\begin{table}[h]
		\centering \caption{\protect\small \textit{
Selected contingency tables from \citet{Efron:1996}.
Posterior  model probabilities,  as a function of $q$ for different comparison scenarios}}
\label{tab:resEfron-PP}
		\begin{tabular}{|c|c||c|c||c|c||c|c|c|}
			\hline
				Table &	&  \multicolumn{2}{|c||}{$M_0$ - $M_e$} & \multicolumn{2}{|c||}{$M_0$ - $M_c$} & \multicolumn{3}{|c|}{$M_0$ - $M_c$ - $M_e$} \\
			\hline
		 34	&	$q$ & {\scriptsize $\prob^I(M_0 |t, \bm{y})$ }& {\scriptsize$\prob^I(M_e |t, \bm{y})$} & {\scriptsize $\prob^I(M_0 |t, \bm{y})$} & {\scriptsize $\prob^I(M_c |t, \bm{y})$ }& {\scriptsize $\prob^I(M_0 |t, \bm{y})$} & {\scriptsize $\prob^I(M_c |t, \bm{y})$} & {\scriptsize $\prob^I(M_e |t, \bm{y})$}\\
			\hline
			&	0   &  0.215 & 0.785 & 0.216& 0.7837 & 0.121 &	0.438 & 0.441	\\
			&	0.25 &  0.172 & 0.828 & 0.234 & 0.766 & 0.115 & 0.329 & 0.555	\\
			&	0.5  & 0.174 & 0.826 & 0.302 & 0.698 & 0.131 & 0.246 & 0.623		\\
			&	0.75 & 0.185 & 0.815 & 0.379 & 0.620 & 0.151 & 0.184 & 0.664	 \\
			&	1    & 0.198 & 0.802 & 0.470 & 0.529 & 0.170 & 0.139 & 	0.691 \\
			\hline
			 1	&	$q$ & {\scriptsize $\prob^I(M_0 |t, \bm{y})$ }& {\scriptsize$\prob^I(M_e |t, \bm{y})$} & {\scriptsize $\prob^I(M_0 |t, \bm{y})$} & {\scriptsize $\prob^I(M_c |t, \bm{y})$ }& {\scriptsize $\prob^I(M_0 |t, \bm{y})$} & {\scriptsize $\prob^I(M_c |t, \bm{y})$} & {\scriptsize $\prob^I(M_e |t, \bm{y})$}\\
			\hline
			& 0 & 0.170 & 0.830 & 0.170 & 0.829 & 0.0929 & 0.4528 &	0.454 \\
			& 0.25 & 0.200 & 0.800 & 0.313 & 0.687 & 0.138 & 0.309 & 0.553 \\
			& 0.5  &  0.225 & 0.775 & 0.454 & 0.546 & 0.174 & 0.227 & 0.598   \\
			& 0.75 & 0.258 & 0.742 & 0.574 & 0.426 & 0.199 & 0.174 & 0.627 \\
			& 1 & 0.258 & 0.742 & 0.258 & 0.742 & 0.223 & 0.134 & 0.643 \\
			\hline
			38	&	$q$ & {\scriptsize $\prob^I(M_0 |t, \bm{y})$ }& {\scriptsize$\prob^I(M_e |t, \bm{y})$} & {\scriptsize $\prob^I(M_0 |t, \bm{y})$} & {\scriptsize $\prob^I(M_c |t, \bm{y})$ }& {\scriptsize $\prob^I(M_0 |t, \bm{y})$} & {\scriptsize $\prob^I(M_c |t, \bm{y})$} & {\scriptsize $\prob^I(M_e |t, \bm{y})$}\\
			\hline
			&0   & 0.395 & 0.605 & 0.398 & 0.601 & 0.248 & 0.374 & 0.379\\
			&0.25 & 0.328 & 0.672 & 0.481 & 0.519 & 0.248 & 0.243 & 0.509 \\
			&0.5  & 0.326 & 0.674 & 0.554 & 0.446 & 0.266 & 0.185 & 0.549 \\
			&0.75 & 0.333 & 0.667 & 0.617 & 0.382 & 0.285 & 0.145 & 0.569 \\
			&	1 & 0.343 & 0.657 & 0.683 & 0.317 & 0.305 & 0.112 & 0.583 \\
			\hline
			18	&	$q$ & {\scriptsize $\prob^I(M_0 |t, \bm{y})$ }& {\scriptsize$\prob^I(M_e |t, \bm{y})$} & {\scriptsize $\prob^I(M_0 |t, \bm{y})$} & {\scriptsize $\prob^I(M_c |t, \bm{y})$ }& {\scriptsize $\prob^I(M_0 |t, \bm{y})$} & {\scriptsize $\prob^I(M_c |t, \bm{y})$} & {\scriptsize $\prob^I(M_e |t, \bm{y})$}\\
			\hline
			&	0   & 0.551 & 0.449 & 0.549 & 0.451 &0.379 & 0.311 & 0.309 \\
			&	0.25 & 0.434 & 0.566 & 0.495 & 0.505 & 0.313 & 0.279 & 0.408\\
			&0.5 & 0.410 & 0.590 & 0.518 & 0.482 & 0.316 & 0.230 & 	0.454\\
			&0.75 & 0.406 & 0.594 & 0.539 & 0.461 & 0.325 & 0.2014 & 0.474 \\
			&1   &  0.410 & 0.590 & 0.408 & 0.592 & 0.340 & 0.172 & 0.489\\
			\hline
			 16	&	$q$ & {\scriptsize $\prob^I(M_0 |t, \bm{y})$ }& {\scriptsize$\prob^I(M_e |t, \bm{y})$} & {\scriptsize $\prob^I(M_0 |t, \bm{y})$} & {\scriptsize $\prob^I(M_c |t, \bm{y})$ }& {\scriptsize $\prob^I(M_0 |t, \bm{y})$} & {\scriptsize $\prob^I(M_c |t, \bm{y})$} & {\scriptsize $\prob^I(M_e |t, \bm{y})$}\\
			\hline
			&	0   & 0.451 & 0.549 & 0.451 & 0.549 & 0.291 & 0.354 & 0.355 \\
			&	0.25 & 0.483 & 0.517 & 0.553 & 0.447 & 0.346 & 0.284 & 0.370\\
			&	0.5  & 0.501 & 0.499 & 0.605 & 0.395 & 0.374 & 0.252 & 0.373\\
			&	0.75 & 0.504 & 0.496 & 0.650 & 0.349 & 0.389 & 0.226 & 0.384\\
			&	1    & 0.498 & 0.502 & 0.674 & 0.326 & 0.396 & 0.208 & 0.396\\
			\hline		
		\end{tabular}
	\end{table}
}
%

%

First  notice that the values of $\prob^I(M_0 |t=0, \bm{y})$ in the $\{ M_0, M_e \}$ scenario coincide with those of in Table 2 of \citet{Case:More:2009} under their \lq \lq Uniform\rq \rq{}
column because in  that case the priors used in the calculation are the default priors \eqref{casella-moreno-priors}. (As usual, we index our table with   $q=t/n$, so that the case $t=0$ coincides with $q=0$.  It can be checked that the results for $t=0$ also coincide with those of $t=1$ because of the structure of formula   \eqref{eq:mult-BF^Ie0}).

%
%
%
%
%
%
%
%

The   results of Table \ref{tab:resEfron-PP}
 can be summarized as follows:

\begin{itemize}
  \item

  For hospitals 34 and 1 there is robust evidence against $M_0$ because its posterior probability is always well below the 50\% threshold, whatever scenario is considered. When all three models are entertained,  it appears that the unconstrained model $M_e$ is better supported than $M_c$, because its posterior probability exceeds 50\% save for $q=0$ (the default starting case which is not recommended for model comparison).

  \item
  For hospitals 18 and 16 the situation is rather  different. The null model of independence achieves values of posterior probability around 40\% when compared against $M_e$; but this probability increases and exceeds 1/2 when the comparison is against $M_c$: this is especially true for hospital 16.
  When all three models are considered jointly it appears that the  unconstrained model prevails although by a moderate amount; interestingly the next better supported model is that of independence,  while the constrained model typically scores the least value.

  \item
  Finally,  for hospital 38 the null model of independence is less supported than for hospitals 18 and 16. In the $\{  M_0, M_e\}$ scenario there is robust but moderate evidence for $M_e$ (the posterior probability being always greater than 0.6); however this is not the case in the
  scenario $\{  M_0, M_c\}$ where $M_0$ receives higher evidence than $M_c$ save for $q=0$ and $q=0.25$.

  This  item is interesting because it reveals that,  by focussing our testing procedure more narrowly on the constrained model (a more reasonable alternative in the context of medical treatment),  the null hypothesis comes out as being more supported by the data.
\end{itemize}

\paragraph
{Self-perception and behavior of students.}

\citet{Nash:Bowen:2002}
considered the relation between perception of internal strength and resources (\lq \lq Internal Assets\rq \rq{}) and class behavior among  students in grades from 6 to 12.
Data were collected using the School Success Profile, a self administered instrument designed for students.
The Internal Assets Index is a measure of  the adolescent’s perception of his
or her strength and resources (health, exercises, or involvement in
sports), which for this study was categorized into \lq \lq low\rq \rq{} and \lq \lq high\rq \rq{}.
Each student was also asked whether he or she
 during the previous
30 days  had been sent away from class because of his or her
behavior.
The data are displayed in Table \ref{tab:tabK&H} where one can verify that the frequency of being sent away from class in the \lq \lq low\rq \rq{} group is only moderately larger than in the \lq \lq high\rq \rq{} group ($0.172>0.136$).

{\renewcommand{\baselinestretch}{1}
\begin{table}[!ht]
\centering \caption{\protect\small \textit{Contingency table $2\times 2$ taken from Table 3 of \citet{Klugkistetal:2010}}} \label{tab:tabK&H}
\begin{tabular}{|c|c|c|}
\hline
Internal Assets  & \multicolumn{2}{|c|}{Sent Away from Class} \\
 & yes & no \\
\hline
low  & 220 & 1060  \\
high & 96  & 609 \\
\hline
\end{tabular}
\end{table}
}
 These data were analyzed by \citet{Klugkistetal:2010} by testing a constrained  model  $M_c$
 (students with low internal assets are more likely to be sent away from class than students with high internal assets)  against the null model $M_0$ of no difference between the two groups; they also considered an unconstrained model $M_e$ as a possible explanation. (Labels for the hypotheses are consistent with the notation in this paper but different from theirs).
 Models $M_0$ and  $M_e$ are defined    as in \eqref{eq:multinomial-M0-M1}, and  default priors
 as in \eqref{casella-moreno-priors}.

 On the other hand model $M_c$ is now specified as
 \be
M_c: \frac{\pi_{11}}{\pi_{11}+\pi_{12}}>\frac{\pi_{21}}{\pi_{21}+\pi_{22}}, \ee


\citet{Klugkistetal:2010} assign only one prior under $M_e$, namely $Dir(\bf{\pi}  | 1,1,1,1)$ and regard both $M_0$ and $M_e$ as constrained models; accordingly they induce priors under the two latter models using an encompassing approach. This however
presents a technical difficulty
because $M_0$ involves only equality constraints which cannot be dealt with directly because the null parameter space is a set of probability zero under the above prior. This leads them to define \lq \lq about equality\rq \rq{} constraints to accommodate the analysis of this problem  into their framework.
 As a consequence they are forced to  introduce  \emph{thresholds} in order to define \lq \lq about equality\rq \rq{}, thus adding an arbitrary step.
  On the other hand our method treats $M_0$ as a separate model with its own parameter and prior.

%
%
%
%
%
%
%
%
%
%
%
%
%

Their analysis leads to the following Bayes factors
 $BF_{ce}(\bm{y})=1.97$ and  $BF_{0e}(\bm{y})=2.59$ which shows
 that the constrained model is better supported than the unconstrained model, however it is the null model which receives higher support.
  The corresponding model posterior
probabilties are $\prob(M_0 | {\bf{y}})=0.47$, $\prob(M_c | {\bf{y}})=0.35$, $\prob(M_e | {\bf{y}})=0.18$, so that the null model of independence appears as the most likely followed by that which assumes a constraint and finally by the unconstrained model.

%
%
%
\black

%

{\renewcommand{\baselinestretch}{0.9}
	\begin{table}[h]
		\centering \caption{\protect\small \textit{Self-perception and behavior of students. Bayes  factors
appearing in formula (\ref{eq:BF^Ic0})
as a function of $q$.
					}} \label{tab:resNashBowen-BF}
		\begin{tabular}{|c|c|c|c|}
			\hline
			$q$ & $BF^I_{e0}$ & $BF^I_{ce}$ & $BF^I_{c0}$  \\
			\hline
			0    & 0.4199 & 1.5386 & 0.6461 \\
			0.25 & 0.4237 & 1.5375 & 0.6515  \\
			0.5  & 0.3864 &	1.6957 & 0.6552    \\
			0.75 & 0.3870 &	1.7893 & 0.6925   \\
			1    & 0.4035 &	1.8531 & 0.7477  \\
			\hline
		\end{tabular}
	\end{table}
}

We report in  Table \ref{tab:resNashBowen-BF} the Bayes factors generated by  our approach. The values of $BF^I_{e0}$ suggest that evidence in favor
of $M_e$ is less than half that  of $M_0$; additionally evidence for $M_c$ is about one-and-half that for $M_e$; as a consequence the evidence for the constrained model
$BF^I_{c0}=BF^I_{e0} \cdot BF^I_{ce}$ ranges between 0.65 and 0.75.
We highlight the fact that there is a good degree of robustness wrt $q$.

In Table \ref{tab:resNashBowen-PP} we report the posterior model probabilities for three distinct comparison scenarios.

{\renewcommand{\baselinestretch}{0.9}
	\begin{table}[h]
		\centering \caption{\protect\small \textit{Self-perception and behavior of students. Posterior model probabilities as a function of $q$ for different  model comparison scenarios }} \label{tab:resNashBowen-PP}
		\begin{tabular}{|c|c|c|c|c|c|c|c|c|}
			\hline
			&  \multicolumn{2}{|c|}{$M_0$ - $M_e$} & \multicolumn{2}{|c|}{$M_0$ - $M_c$} & \multicolumn{3}{|c|}{$M_0$ - $M_c$ - $M_e$} \\
			\hline
			$q$ & {\scriptsize $\prob^I(M_0 |t, \bm{y})$ }& {\scriptsize$\prob^I(M_e |t, \bm{y})$} & {\scriptsize $\prob^I(M_0 |t, \bm{y})$} & {\scriptsize $\prob^I(M_c |t, \bm{y})$ }& {\scriptsize $\prob^I(M_0 |t, \bm{y})$} & {\scriptsize $\prob^I(M_c |t, \bm{y})$} & {\scriptsize $\prob^I(M_e |t, \bm{y})$}\\
			\hline
			0    & 0.7043 &	0.2957 & 0.6075 & 0.3925 & 0.4840 & 0.3127 & 0.2032 \\
			0.25 & 0.7024 &	0.2976 & 0.6055 & 0.3945 & 0.4819 & 0.3139 & 0.2042 \\
			0.5  & 0.7213 &	0.2787 & 0.6042 & 0.3958 & 0.4898 & 0.3209 & 0.1893 \\
			0.75 & 0.7210 & 0.2790	 & 0.5904 & 0.4096 & 0.4808 & 0.3330 & 0.1861 \\
			1    & 0.7125 & 0.2875 & 0.5722 & 0.4278 & 0.4649 & 0.3476 & 0.1876  \\
			\hline
		\end{tabular}
	\end{table}
}

It appears from Table \ref{tab:resNashBowen-PP} that the null model of independence should be regarded as the one having the strongest evidential support, whatever the scenario under consideration.
We believe that the most reasonable comparison setting for testing independence in this problem is the one involving $\{ M_0, M_c \}$ because the natural expectation is that  students with low internal assets are more likely, if at all,  to be sent away from class than students with high internal assets. This is also somewhat supported on purely descriptive  grounds, as already recalled.
In the above scenario,  the posterior probability of $M_0$ ranges between 57 and 61\% as $q$ varies; in other words there is evidence in favor of $H_0$ because the conventional $50\%$ threshold is exceeded, although its strength is not overwhelming.
Interestingly when $M_0$ is contrasted with the unconstrained model $M_e$ its evidence increases to a comfortable 70\%, and again this result is robust. The reason is clear: the alternative confronting $M_0$ is now less precise (the parameter space is larger)  and,  more importantly,  in greater conflict with the likelihood than $M_c$ (recall that the frequency of being Sent Away from Class with low Internal assets is 0.172 whereas the corresponding frequency with high Internal Assets  is 0.136, a lower value). As a consequence,   its marginal likelihood is penalized and this is reflected in a lower (higher) posterior probability for $M_e$ ($M_0)$.
Finally, for the scenario involving three models $\{ M_0, M_c, M_e \}$,  $M_0$ remains  the highest-posterior-probability model although its value drops to around 47-48\%. This is only to be expected because we are now contrasting $M_0$ with two, rather than just one, competitors.
We could view this  comparative scenario as a sort of \lq \lq stress test\rq \rq{} for $M_0$, rather than a substantive research hypothesis. Since  the threshold $50\%$ is only closely missed, evidence for $H_0$ seems worth of consideration.
\section{Concluding remarks}

\label{sec:concluding}
In this paper we have presented an objective Bayes model comparison procedure for two-way contingency tables where models are specified by inequality constraints on the parameter space.
Specifically we have considered contingency tables whose sampling distribution is either  product multinomial (including product Binomial as a special case),  or fully multinomial.

Our method relies on the principled method of  intrinsic priors for model comparison coupled
with the encompassing prior approach to compute the Bayes factor and the posterior probability of each constrained model.

An attractive  feature of our method
is that it requires no subjective input, but merely standard default priors, which makes it essentially fully automatic.
When the default prior, as in our examples, is already proper,  we can assess the robustness of our inference by letting the fraction of prior sample size to actual sample size vary: if the result is always above a threshold deemed to represent sufficient evidence,  then we can safely conclude that the result is robust.
Our method can also deal with starting default improper priors however, such as the Jeffreys prior $\mbox{Beta}(1/2,1/2)$: in this case however we would recommend using \emph{conditionally intrinsic} priors, as opposed to fully intrinsic priors: for details see \citet{Cons:Paro:2017}.

Another interesting characteristic of our method is that it can deal simultaneously with nested models having different or equal dimensions of the parameter space. For instance,  the constrained  models in  the examples of Section 5 were all of the same dimension as the unconstrained model.
Since the Bayes factor has a built-in   Ockam's razor,
 it can well happen that the constrained model receives higher evidence than the encompassing model because it trades automatically   fit and model complexity.

  Using the  general principles expounded in \citet{Cons:Paro:2017},  
 the scope of our  method  can be extended  to the  comparison of    models specified by equality \textit{and} inequality constraints. This can be done exactly   without resorting to  approximate representations of equality constraints.  

\bibliographystyle{biometrika}
\bibliography{Constrained}

\newpage

\section*{Appendix}
\subsection*{Bayes Factor in the multinomial case}

\subsubsection*{A1. Algorithm to estimate $BF^I_{e0}(\bm{y}|t)$ }
To estimate ${BF}^{I}_{e0}(\bm{y}|t)$ we use an importance sampling  algorithm to compute the MonteCarlo sum.

\begin{algorithm}
	\caption{Approximation of the  ${BF}^{I}_{e0}(\bm{y}|t)$  }\label{alg:BFe0}
	\begin{algorithmic}[1]
		\For{$s=1,2,\ldots,S$}
		\State sample a contingency table $\bm{x}^{(s)}$ from the candidate distribution that is  Multinomial with cell probabilities equal to the observed cell relative frequencies and with total sum equal to $t$:
		\be
		\bm{x}^{(s)}\sim Multinomial(t,\bm{\hat{\pi}})
		\ee
		with $\bm{\hat{\pi}}=\{ \hat{\pi}_{ij}=y_{ij}/n; i=1,\ldots,r; j=1,\ldots,c \}$;
		 \State compute the product
		 \be
		 q^{(s)}= \frac{(\prod_{i}x_{i+}^{(s)}!)(\prod_{j}x_{+j}^{(s)}!)}{\prod_{i}y_{i+}!)(\prod_{j}y_{+j}!)}
		 \times \frac{\prod_{i,j}(x_{ij}^{(s)}+y_{ij})!}{\prod_{i,j}x_{ij}^{(s)}!} \times \frac{1}{\prod_{ij}\hat{\pi}_{ij}^{x_{ij}^{(s)}}}
		 \ee
		 where the last element is the importance density of the sampled table;
		\EndFor
		\State \textbf{end for}
		\State approximate ${BF}^{I}_{e0}(\bm{y}|t)$ as
		\be
		\widehat{BF}^{I}_{e0}(\bm{y}|t)\simeq \frac{\Gamma(t+rc)\Gamma(n+r)\Gamma(n+c)}{\Gamma(t+n+rc)\Gamma(t+r)\Gamma(t+c)}
		\times \frac{1}{S} \sum_{s=1}^{S} q^{(s)}.
		\ee
	\end{algorithmic}
\end{algorithm}

%
%
\subsubsection*{A2. Algorithms to estimate  $BF^I_{ce}(\bm{y}|t)$}
 To compute ${BF}^{I}_{ce}(\bm{y}|t)$ we use the same strategy discussed in par. (\ref{sec:BFce})
\bitem
\item [(a)]
   Consider first the evaluation of the \textit{denominator}. We approximate the required probability by drawing $S$ samples from the intrinsic prior \eqref{priorBFce}, check the constraints and compute the fraction of samples that satisfy them.

\begin{algorithm}
	\caption{Approximation of the denominator of ${BF}^{I}_{ce}(\bm{y}|t)$  }\label{alg:denBFce}
	\begin{algorithmic}[1]
		\For{$s=1,2,\ldots,S$}
		\State sample $\bm{\pi}^{*(s)}$ under the null model; i.e. sample the marginals $\bm{\pi_1}^{(s)}$ and $\bm{\pi_2}^{(s)}$ from their priors
		$\bm{\pi_1}^{(s)}\sim Dir(\bm{\alpha}_{\pi_1})$ and $\bm{\pi_2}^{(s)}\sim Dir(\bm{\alpha}_{\pi_2})$, and compute $\bm{\pi}^{*(s)}=\{\pi_{ij}^{*(s)}=\pi_{1i}^{(s)}\cdot \pi_{2j}^{(s)}; i=1.\ldots,r; j=1,\ldots,c \}$;
		\State sample a contingency table $\bm{x}^{(s)}$ from a Multinomial:
		\be
		\bm{x}^{(s)} \sim  Multinomial(t,\bm{\pi}^{*(s)});
		\ee
		\State sample $\bm{\pi}^{(s)}$ from the pseudo posterior
		\be
		\bm{\pi}^{(s)} \sim Dir(\bm{\alpha}_{\pi}+\bm{x}^{(s)});
		\ee
		\EndFor
		\State \textbf{end for}
		\State approximate the probability in the denominator as
		\be
		\widehat{\prob}^{I} \{ \bm{\pi}\in \Theta_c |M_e \} \thickapprox\frac{1}{S}\sum_{s=1}^{S} \bm{1}_{\bm{\pi}\in \Theta_c}(\bm{\pi}^{(s)}).
		\ee
		\end{algorithmic}
\end{algorithm}

\item [(b)]
For the  \textit{numerator} of $BF^I_{ce}$ we rely on a Metropolis within Gibbs algorithm as follows:

 \begin{algorithm}
 	\caption{Approximation of the numerator of ${BF}^{I}_{ce}(\bm{y}|t)$  }\label{alg:enumBFce}
 	\begin{algorithmic}[1]
 		\State  estimate the probabilities $\bm{\pi}$ with the modified MLE estimators
 		\be
 		\widehat{\pi}_{ij}=\frac{y_{ij}+1}{n+rc}, \ \ \ \forall i=1,\ldots,r \ and\  j=1,\ldots,c;
 		\ee
 		\State sample  starting table from a Multinomial with parameters $\widehat{\bm{\pi}}$ and sum equal to $t$:
 		$\bm{x}^{(0)}\sim Multinomial(t,\widehat{\bm{\pi}})$;
 		\For{$s=1,2,\ldots,S_1$}
 		\State generate  $\bm{x}^{(s)}$ by sampling from the proposal $Multinomial(t,\widehat{\bm{\pi}})$ and accepting probability:
 		\be
 		\alpha\left(\bm{x}^{(s-1)};\bm{x}^{(s)}\right) =\min \left\{ 1;%
 		\frac{H(\bm{x}^{(s)}|t,\bm{y}) Multinonial(\bm{x}^{(s-1)}|t,\widehat{\bm{\pi}})
 		}{H(\bm{x}^{(s-1)}| t,\bm{y})
 			Multinonial(\bm{x}^{(s)}|t,\widehat{\bm{\pi}})  }
 		\right\};
 		\ee
 		where $H(\bm{x}| t,\bm{y})$ is:
 		\be
 		H(\bm{x} |t,\bm{y}) &=&
 		\binom{n}{\bm{y}} \frac{B(\bm{\alpha}_{\pi}+\bm{y}+\bm{x})}{B(\bm{\alpha}_{\pi}+\bm{x})}
 		\binom{t}{\bm{x}} \frac{B(\bm{\alpha}_{\pi_1}+\bm{x_{R+}})}{B(\bm{\alpha}_{\pi_2}+\bm{x_{+C}})};
 		\ee
 		\EndFor  \State \textbf{end for}
 		\For{$s=S_1+1,\ldots,S_1+S$}
 		\State repeat step (4);
 		\State sample each component of  $\bm{\pi}^{(s)}$ from
 		$\bm{\pi}^{(s)}|t, \bm{x}^{(s)},M_e \sim Di(\bm{\alpha}_{\pi}+\bm{x^{(s)}})$;
 		\EndFor 	\State \textbf{end for}
 		\State approximate the probability of the numerator  as
 		\be \widehat{\prob}^{I} \{ \bm{\pi}\in \Theta_c |\bm{y},t,M_e \} \thickapprox\frac{1}{S}\sum_{s=S_1+1}^{S_1+S} \bm{1}_{\bm{\pi}\in \Theta_c }(\bm{\pi}^{(s)}).
 		\ee
 		\end{algorithmic}
 \end{algorithm}

\item [(c)]
Finally obtain
\be
\widehat{BF}^{I}_{c,e}(\bm{y}|t)=
\frac{\widehat{\prob}^{I} \{\bm{\pi}\in \Theta_c |\bm{y},t,M_e  \}}
{\widehat{\prob}^{I} \{ \bm{\pi}\in \Theta_c|t, M_e  \}}.
\ee

\eitem


\end{document}